\pdfoutput=1
\documentclass[a4paper,11pt]{article}
\usepackage[utf8]{inputenc}
\usepackage{jcappub}
\usepackage[utf8]{inputenc}
\usepackage{amsmath,amssymb}
\usepackage{graphicx}
\usepackage{xcolor}
\usepackage{multicol}
\usepackage{hyperref}
\usepackage{physics}
\usepackage{lipsum}
\usepackage{caption}
\usepackage{subcaption}
\usepackage{accents}
\usepackage{caption}
\usepackage{subcaption}
\usepackage{dsfont}
\usepackage{physics}
\usepackage{mathtools,leftindex}
\usepackage{multirow}
\usepackage{comment}
\usepackage{array}
\usepackage{float}

\newcommand{\m}{m_{\text{P}}}

\usepackage{pgfplots}
\usepgfplotslibrary{fillbetween}
\usetikzlibrary{patterns}
\pgfplotsset{compat=1.8}

\newcommand{\be}{\begin{equation}} 
\newcommand{\ee}{\end{equation}}
\newcommand{\bea}{\begin{equation}\begin{aligned}} 
\newcommand{\eea}{\end{aligned}\end{equation}}
\newcommand{\ba}{\begin{eqnarray}}
\newcommand{\ea}{\end{eqnarray}}

\title{\boldmath Non-Minimally Coupled Quintessence in Light of DESI}
\author[a,b]{Samuel S\'anchez L\'opez,}
\author[c]{Alexandros Karam,}
\author[a,b,d]{Dhiraj Kumar Hazra}

\emailAdd{ssanchezlopez@imsc.res.in} 
\emailAdd{alexandros.karam@kbfi.ee} 
\emailAdd{dhiraj@imsc.res.in}

\vspace{5cm}
\affiliation[a]{The Institute of Mathematical Sciences, HBNI, CIT Campus, Chennai 600113, India}
 \affiliation[b]{Homi Bhabha National Institute, Training School Complex, Anushakti Nagar, Mumbai 400094, India}
\affiliation[c]{National Institute of Chemical Physics and Biophysics, Rävala 10, 10143 Tallinn, Estonia}
 \affiliation[d]{INAF/OAS Bologna, Osservatorio di Astrofisica e Scienza dello Spazio, Area della ricerca CNR-INAF, via Gobetti 101, I-40129 Bologna, Italy}

\abstract{We analyze a model of quintessence governed by an exponential potential and non-minimally coupled to gravity, in light of recent datasets, including cosmic microwave background, baryon acoustic oscillations, and supernovae distance moduli observations. Mainly focusing on the Palatini formulation of gravity, a phase space analysis reveals the existence of a late-time stable de Sitter attractor as long as the non-minimal coupling constant is negative, regardless of the value of the slope of the exponential. Fitting to CMB+DESI+DESY5 data, we find strong evidence for our model over $\Lambda$CDM, with a Bayes factor $\log B = 5.52$. Furthermore, the data seem to prefer dynamical dark energy at $>3\sigma$ C.L. and a phantom crossing in the barotropic parameter of dark energy at $2-3\sigma$ C.L.. We find that the scalar field dynamics in the Palatini formalism provides marginally better agreement to the data compared to the metric formalism.}

\begin{document}

\maketitle
\section{Introduction}

The entire history of the Universe, from the initial conditions of the Hot Big Bang, provided by primordial quantum fluctuations stretched to super-horizon scales~\cite{Starobinsky:1980te, Guth:1980zm, Linde:1981mu, Albrecht:1982wi}, until the present day, 13.8 billion years later, is well described by a six-parameter model dubbed $\Lambda$CDM with a power law form of primordial spectrum, or the standard model of cosmology. Two of the six parameters correspond to the amplitude and tilt of a nearly scale-invariant spectrum for the initial conditions, while the remaining four are linked to background quantities, namely the density parameters of baryonic and cold dark matter (CDM), the reionization depth, and the angular size of the horizon at recombination. $\Lambda$CDM, which serves as a baseline model, not only is in excellent agreement with the most precise Cosmic Microwave Background (CMB) data from Planck~\cite{Planck:2018vyg}, but also successfully addresses the low redshift observations, making it the most successful cosmological model to date. 

Despite this triumphant success, a more careful analysis reveals tensions between cosmological observations when the standard model is used to simultaneously fit high- and low-redshift datasets in a joint analysis. For example, the value of the Hubble parameter inferred from CMB data is about 8\% smaller, at a confidence level of $5\sigma$, than the value locally measured by using the distance ladder~\cite{Riess:2021jrx}, in what is called the Hubble tension~\cite{Kamionkowski:2022pkx, Bernal:2016gxb}. Other tensions include discrepancies in the inferred value of the matter clustering parameter $S_8$ or the CMB lensing amplitude $A_{\rm lens}$ anomaly (see Ref.~\cite{Abdalla:2022yfr} for a comprehensive review). More recently, baryon acoustic oscillations (BAO) measurements~\cite{DESI:2024mwx, DESI:2025zgx} reveal mild tensions with CMB data, particularly when combined with type Ia supernovae (SNIa) observations. The focus of the present work will be the latter, by considering a model of dynamical dark energy featuring deviations from a cosmological constant behaviour at late times, something that has been the subject of intense study in the recent past~\cite{Khoury:2025txd, Pan:2025psn, Nesseris:2025lke, Pan:2025qwy, Kessler:2025kju, Tiwari:2024gzo, Berbig:2024aee, Shlivko:2025fgv, Benisty:2025mfv, Akrami:2025zlb, Wolf:2025jed, Mirpoorian:2025rfp, Wang:2025bkk, Dinda:2025iaq, wang2025questioningcosmicaccelerationdesi, cortês2025desisdr2exclusionlambdacdm, deSouza:2025rhv, Afroz:2025iwo, Garcia-Quintero:2025qeq, Scherer:2025esj, Chen:2025mlf, Ye:2025ark, Chen:2025wwn, Smirnov:2025yru, Giani:2025hhs, Wang:2025rll, Hussain:2025nqy, Wang:2025vfb, Lee:2025hjw, Colgain:2025fct, Bayat:2025xfr, DAgostino:2025kme, Sabogal:2025jbo, Myrzakulov:2025jpk, Cline:2025sbt, Keeley:2025rlg, Anchordoqui:2025epz, Lee:2025ysg, Lee:2025kbn, Ozulker:2025ehg, Gialamas:2025pwv, Li:2025ops, Dhawan:2025mer, Lee:2025rmg, Liu:2025myr, Hogas:2025ahb, Qiang:2025cxp, Mishra:2025goj, Bhattacharjee:2025xeb, Braglia:2025gdo, Goldstein:2025epp, Chaudhary:2025pcc, Wang:2025vtw, Ishak:2025cay, Torres:2025qko, Gomez-Valent:2025mfl, Wang:2025znm, Chen:2025ywv, Goh:2025upc, Wolf:2025acj, Luciano:2025ykr, Chaudhary:2025uzr, RoyChoudhury:2025iis, Park:2025fbl, Artola:2025zzb, Guedezounme:2025wav, Chaudhary:2025bfs, Li:2025muv, Rezaei:2025vhb, Zhang:2025dwu, Toomey:2025xyo, Yao:2025twv}.

In this paper, we take a theoretically motivated approach, keeping simplicity in mind. Thus, the role of dynamical dark energy is played by a scalar field $\phi$. Scalar fields are ubiquitous in high energy physics, including the Higgs mechanism~\cite{Englert:1964et, Higgs:1964ia}, inflation~\cite{Martin:2013tda}, dark matter~\cite{Hu:2000ke, Cline:2013gha, Marsh:2015xka}, string theory~\cite{Svrcek:2006yi, Cicoli:2012sz, Arvanitaki:2009fg}, modified gravity~\cite{Faraoni:2001tq, DeFelice:2010aj}, and, in our case, quintessence~\cite{Ratra:1987rm, Wetterich:1987fm, Caldwell:1997ii}. Furthermore, we include a non-minimal coupling to gravity in the action governing its dynamics. Indeed, in the context of quantum field theory in curved spacetime, even if at tree level the field is minimally coupled, renormalization requires the inclusion of counterterms that couple it to the Ricci scalar $R$. In this way, the action at loop level must include a term proportional to $\phi^2 R$~\cite{Birrell:1982ix, Parker:2009uva, Callan:1970ze}. 

The inclusion of a non-minimal coupling in the action makes the dynamics sensitive to the formalism of the theory of gravity. In this work we mainly focus on the Palatini formalism \cite{Palatini:1919ffw, Ferraris:1982wci}, which has gained decisive momentum in recent years \cite{Bauer:2008zj,Bauer:2010bu,Tamanini:2010uq, Bauer:2010jg, Rasanen:2017ivk, Tenkanen:2017jih, Racioppi:2017spw, Markkanen:2017tun, Jarv:2017azx, Fu:2017iqg, Racioppi:2018zoy, Carrilho:2018ffi, Kozak:2018vlp, Rasanen:2018fom, Rasanen:2018ihz, Almeida:2018oid, Shimada:2018lnm, Takahashi:2018brt, Jinno:2018jei, Rubio:2019ypq, Bostan:2019uvv, Bostan:2019wsd, Racioppi:2019jsp, Tenkanen:2020dge, Shaposhnikov:2020fdv, Borowiec:2020lfx, Jarv:2020qqm, Karam:2020rpa, McDonald:2020lpz, Langvik:2020nrs,Tenkanen:2019xzn, Shaposhnikov:2020gts, Shaposhnikov:2020frq, Gialamas:2020vto, Mikura:2020qhc, Verner:2020gfa, Enckell:2020lvn, Reyimuaji:2020goi, Karam:2021wzz, Mikura:2021ldx, Kubota:2020ehu, Gomez:2021roj, Mikura:2021clt, Bombacigno:2018tyw, Enckell:2018hmo, Antoniadis:2018ywb, Antoniadis:2018yfq, Tenkanen:2019jiq, Edery:2019txq, Giovannini:2019mgk, Tenkanen:2019wsd, Gialamas:2019nly, Tenkanen:2020cvw, Lloyd-Stubbs:2020pvx, Antoniadis:2020dfq, Ghilencea:2020piz,Das:2020kff, Gialamas:2020snr, Ghilencea:2020rxc, Bekov:2020dww, Gomez:2020rnq, Dimopoulos:2020pas, Karam:2021sno, Lykkas:2021vax, Gialamas:2021enw, Annala:2021zdt, AlHallak:2021hwb, Dioguardi:2021fmr, Dimopoulos:2022tvn, Gialamas:2023flv, SanchezLopez:2023ixx, TerenteDiaz:2023kgc, Kuralkar:2025hoz, Lopez:2025gfu}, and emphasize the differences with respect to the widely used metric formalism throughout the paper. In the Palatini formalism, the connection is taken to be \textit{a priori} independent in such a way that the action should be varied with respect to it, in addition to the metric. The result is that the field equations acquire additional terms relative to their metric counterparts, in which the connection is fixed to the Levi-Civita form, leading to different dynamics. This relatively subtle point is non-existent for an Einstein-Hilbert action. In this case, the variation of the action with respect to the connection dynamically fixes it to its Levi-Civita form, and both the metric and Palatini formalisms agree.

In this work, we analyze for the first time a model of Palatini non-minimally coupled quintessence in the light of state-of-the-art cosmological data, including CMB~\cite{Planck:2018vyg}, BAO~\cite{DESI:2025zpo, DESI:2025zgx}, and SNe~\cite{DES:2024jxu} observations. Our focus is both theoretical, performing a phase space analysis of the model, and observational, including a thorough statistical study of our results. Our work emphasizes the capability of current data to probe modifications to general relativity, as well as the degrees of freedom of the theory of gravity itself.

The paper is organized as follows. In Sec.~\ref{sec:model} we lay out the theoretical aspects of the model. In Sec.~\ref{asdifbaiuhdfdwerr} we compute the field equations, explicitly showing the difference between the metric and Palatini formalisms, and in Sec.~\ref{sec:dynamicalsystem} we write the field equations as an autonomous dynamical system, providing the phase space analysis. Sec.~\ref{asodfjbiawerwer} is devoted to describing the datasets and methodology used to constrain our model, and we present the results in Sec.~\ref{sec:results}. In Sec.~\ref{easdfbahuisdfawew} we show the improvement in fit to different datasets and the resulting parameter posteriors, and in Sec.~\ref{eqaouosdbfadfdawer} we compare the fits to the data of both the Palatini and metric formalisms. Finally, in Sec.~\ref{ref:conclusions} we give our concluding remarks and outlook.

Greek indices represent space-time coordinates $\mu,\nu=0,1,2,3$ and Latin indices represent spatial coordinates $i,j=1,2,3$. Repeated indices are summed over. We assume natural units with $c=\hbar = 1$ and $\m = 1/\sqrt{8\pi G_{\rm N}}=2.44\cross 10^{18} \ \textrm{GeV}$, where $\m$ is the reduced Planck mass. The signature of the metric is mostly positive $(-,+,+,+)$.

\section{The model}
\label{sec:model}
In this section, we present our model of non-minimally coupled quintessence, which will then be analyzed in the light of different datasets. Previous works on the subject include Refs.~\cite{Wolf:2024stt, Ye:2024ywg, Ye:2024zpk, Ferrari:2025egk}, utilizing the DESI DR1 data, and the more recent Refs.~\cite{Pan:2025psn, Tiwari:2024gzo, Wolf:2025jed, Myrzakulov:2025jpk, Wang:2025znm} utilizing the DESI DR2 data. Even though they study different models, what they all have in common is the underlying formalism of gravity: the metric formalism. In the present work, we mainly focus on the Palatini formalism, although we shall also give important results in the metric formalism, both for the sake of comparison and to emphasize the differences between the two formalisms. To the best of our knowledge, our model was only previously considered in Ref.~\cite{Antoniadis:2022cqh} as a standalone quintessence model and in Refs.~\cite{Dimopoulos:2022tvn, Dimopoulos:2022rdp} in the context of quintessential inflation. 

In what follows, we give a brief overview of the dynamics of a scalar field non-minimally coupled to gravity, deriving the field equations in both formalisms. The interested reader may consult \textit{e.g.}~\cite{Gialamas:2023flv} for further details. We then express the equations of motion as a dynamical system and provide the phase space analysis. This approach was previously considered in Refs.~\cite{Jarv:2021qpp, Jarv:2024krk, Jarv:2025qgo} in the context of inflation and in Refs.~\cite{Antoniadis:2022cqh,Wang:2012rva, Fan:2015rha} in the context of quintessence.

\subsection{Action and field equations} \label{asdifbaiuhdfdwerr}
We consider a 
canonical scalar field, $\phi$, which plays the role of quintessence. It is non-minimally coupled to gravity and minimally coupled to the matter and radiation sectors. The action in the Jordan frame reads
\be 
    S =  \int \dd^4 x \sqrt{-g} \left[ \frac{\m^2}{2}f(\phi) \mathcal{R} - \frac{1}{2} g^{\mu\nu} \partial_\mu \phi \partial_\nu \phi - V(\phi) \right] + S_m \left[ g_{\mu\nu} , \chi_m \right] \,,
    \label{eq:action}
\ee
where $g_{\mu\nu}$ is the metric tensor, $S_m$ is the matter action, $\chi_m$ collectively represents the matter fields, and $\mathcal{R}$ is the Ricci scalar, which is obtained by contracting the metric with the Ricci tensor $\mathcal{R}=g^{\mu\nu}\mathcal{R}_{\mu\nu}$. The latter is obtained from the contraction of the Riemann tensor $\mathcal{R}^{\alpha}_{\phantom{h}\mu\alpha\nu}$ and can be written solely in terms of the connection as
\begin{equation}
    \mathcal{R}_{\mu\nu} = \partial_\lambda \Gamma^\lambda _{\mu\nu}-\partial_\nu \Gamma^\lambda _{\mu\lambda}+\Gamma^\lambda_{\lambda\rho}\Gamma^\rho_{\mu\nu} - \Gamma^\rho_{\mu\lambda}\Gamma^\lambda_{\nu\rho}.
\end{equation}
In the metric formalism of gravity, the connection takes the Levi-Civita form given by
\begin{equation}
    \label{eq:levi-civita-connection-eq}L^{\mu}_{\alpha\beta}=\frac{1}{2}g^{\mu\lambda}\left(\partial_{\alpha}g_{\lambda\beta}+\partial_{\beta}g_{\lambda \alpha}-\partial_{\lambda}g_{\alpha \beta}\right),
\end{equation}
which depends only on the metric. However, \textit{a priori} the connection and the metric need not be related. This is so in the Palatini formalism \cite{Palatini:1919ffw, Ferraris:1982wci}, where the connection is assumed to be an independent gravitational field, denoted as $\hat{\Gamma}^{\mu}_{\alpha\beta}$. Consequently, in order to obtain the equations of motion, one must vary the action with respect to both $g_{\mu\nu}$ and $\hat{\Gamma}^{\mu}_{\alpha\beta}$. Hereinafter, we denote tensors constructed using the independent connection with a hat and tensors constructed using the Levi-Civita connection without a hat. Scripted quantities correspond to either metric or Palatini, as in the action~\eqref{eq:action}.

It is important to mention that in a theory with a pure Einstein–Hilbert (EH) action (and minimally coupled matter), the metric and Palatini variational formalisms are dynamically equivalent. However, once the action is extended, \textit{e.g.}~by a non-minimal coupling $f(\phi)\mathcal{R}$ or by higher-curvature terms such as $F(\mathcal{R})$, the two approaches yield different field equations and, hence, distinct cosmological dynamics.
To understand why, one can take the Palatini action in the non-minimal coupling case, where $\mathcal{R}=g^{\mu\nu}\hat{R}_{\mu\nu}$, and vary it with respect to $\hat{\Gamma}^{\mu}_{\alpha\beta}$, giving
\be 
    \hat{\nabla}_\lambda \left( \sqrt{-g} f g_{\mu\nu} \right) = 0 \,.
\ee
This means that $\hat{\Gamma}^{\lambda}_{\ \mu\nu}$ is compatible with $h_{\mu\nu} = f g_{\mu\nu}$. Therefore,
\be 
\label{eq:aiudbfafawer}
    \hat{\Gamma}^{\mu}_{\alpha\beta} = \frac{1}{2}h^{\mu\lambda}\left(\partial_{\alpha}h_{\lambda\beta}+\partial_{\beta}h_{\lambda \alpha}-\partial_{\lambda}h_{\alpha \beta}\right)= L^{\mu}_{\alpha\beta}+\frac{1}{2} \left[\delta^{\mu}_{\beta}\partial_\alpha \log f+\delta_{\alpha}^{\mu} \partial_{\beta} \log f - g_{\alpha\beta} \partial^{\mu} \log f \right] \,.
\ee
It is clear that for a minimally coupled scalar field $f(\phi)=1$, $\hat{\Gamma}^{\mu}_{\alpha\beta}= L^{\mu}_{\alpha\beta}$. However, for any other function of the field $f(\phi)$, the Ricci tensor $\hat{R}_{\mu\nu}$ will acquire additional terms with respect to $R_{\mu\nu}$, coming from the last bracket in Eq. \eqref{eq:aiudbfafawer} (\textit{e.g.} see Eq. \eqref{asdijbjfiawerwe}).

We take the non-minimal coupling function to be $f(\phi) = 1 + \xi \phi^2/\m^2$, extensively studied in the context of inflation \cite{Bauer:2008zj, Bauer:2011sft, Tamanini:2010uq, Bauer:2010jg, Rasanen:2017ivk, Tenkanen:2017jih, Racioppi:2017spw, Markkanen:2017tun, Jarv:2017azx, Fu:2017iqg, Racioppi:2018zoy, Carrilho:2018ffi, Kozak:2018vlp, Rasanen:2018fom, Rasanen:2018ihz, Almeida:2018oid, Shimada:2018lnm, Takahashi:2018brt, Jinno:2018jei, Rubio:2019ypq, Bostan:2019uvv, Bostan:2019wsd, Racioppi:2019jsp, Tenkanen:2020dge, Shaposhnikov:2020fdv, Borowiec:2020lfx, Jarv:2020qqm, Karam:2020rpa, McDonald:2020lpz, Langvik:2020nrs, Shaposhnikov:2020gts, Shaposhnikov:2020frq, Mikura:2020qhc, Gialamas:2020vto, Verner:2020gfa, Enckell:2020lvn, Reyimuaji:2020goi, Karam:2021wzz, Racioppi:2021ynx, Mikura:2021clt, Cheong:2021kyc, Racioppi:2021jai, Kodama:2021yrm, Karananas:2022byw, Yin:2022fgo, Gialamas:2022gxv, Gialamas:2023flv, Jarv:2024krk}. Indeed, even if a theory is minimally coupled at tree level, a non-minimal coupling of this form will be generated at the loop level~\cite{Callan:1970ze, Birrell:1982ix}. The function $f(\phi)$ rescales the effective Planck mass as $M_{\rm eff}^2(\phi)\equiv m_p^2 f(\phi)$, so one must require $f(\phi)>0$ at all times. In the case of $f(\phi) = 1$ and a fixed scalar field, the action reduces to Einstein gravity with the potential playing the role of a cosmological constant. 

The non-minimal coupling $f(\phi) \mathcal{R}$ allows one to recast the theory by a conformal transformation to the Einstein frame, where the gravitational sector takes the EH form and the scalar field is canonical (after a field redefinition). The trade-off is that the matter sector is no longer minimally coupled to the metric, and the usual conservation law for the matter energy-momentum tensor~\cite{Koivisto:2005yk} does not hold in its standard form. This is the issue faced by the authors of Ref.~\cite{Antoniadis:2022cqh}, where these new couplings are neglected as a simplifying assumption to their analysis in the Einstein frame. In the present work, we choose to work exclusively in the Jordan frame, making our treatment exact. Note that although the two frames are mathematically related, they are not physically equivalent unless one simultaneously adopts variable units in the Einstein frame.

We further adopt an exponential potential for the scalar field, given by
\be \label{eq:potential}
    V = V_0 e^{-\lambda \phi/\m} \,,
\ee
where $\lambda$ is a constant that controls the slope, with $\lambda > 0$, $V_0 \geq 0$. This potential is a minimal, theoretically motivated choice, which commonly appears in string theory and supergravity~\cite{Gorlich:2004qm, Haack:2006cy, Lalak:2005hr} and has been extensively studied in different cosmological scenarios~\cite{Caldwell:1997ii, Copeland:1997et, Rubano:2001su, Barreiro:1999zs, Bhattacharya:2024hep, Alestas:2024gxe, Ramadan:2024kmn, Wetterich:1994bg, Binetruy:1998rz, Agrawal:2018own, Akrami:2018ylq, Raveri:2018ddi}. It provides a clean baseline for assessing the impact of the non-minimal coupling on the background expansion.

Varying the action~\eqref{eq:action} with respect to the metric tensor $g_{\mu\nu}$, we obtain the field equations
\be 
    f \mathcal{R}_{\mu\nu} - \frac{1}{2} f \mathcal{R} g_{\mu\nu} - \left( 1 - \delta_P \right) \left( \nabla_\mu \nabla_\nu f - g_{\mu\nu} \nabla_\sigma \nabla^\sigma f \right) = \frac{1}{\m^2} \left[ T_{\mu\nu}^{(\phi)} + T_{\mu\nu}^{({\rm m,r})} \right] \,,
    \label{eq:field_Eqs}
\ee
where
\begin{align}
\label{eq: metric Palatini delta}
    \delta_{\rm P} = \left\lbrace \begin{tabular}{ll}
        0 \,, & \textrm{metric} \\
        1 \,, & \textrm{Palatini}
    \end{tabular} \right. 
\end{align}
and we emphasize the unhatted $\nabla_\mu$ is the covariant derivative related to the Levi-Civita connection. In Eq.~\eqref{eq:field_Eqs}, the energy-momentum tensor of quintessence reads
\begin{equation}
    T^{(\phi)}_{\mu\nu}=\partial_\mu\phi\,\partial_\nu\phi-\tfrac12 g_{\mu\nu}(\partial\phi)^2-g_{\mu\nu}V(\phi)
\end{equation}
and $T^{({\rm m,r})}_{\mu\nu}$ is the combined matter–radiation tensor, taken to be that of a perfect fluid
\be 
    T^{\rm (m,r)}_{\mu\nu} = - \frac{2}{\sqrt{-g}} \frac{\delta S_m}{\delta g^{\mu\nu}}=(\rho+p)u_\mu u_\nu+pg_{\mu\nu} \,.
\ee
Using the metric Einstein tensor $G^{\mu}_{\ \nu} = R^{\mu}_{\ \nu} - \frac{1}{2} R g^{\mu}_{\ \nu}$, Eq.~\eqref{eq:field_Eqs} can be cast in the standard form
\be 
    G^{\mu}_{\ \nu} = \frac{1}{\m^2} \left[ T^{\mu (\phi)}_{\ \nu} + T^{\mu ({\rm m,r})}_{\ \nu} + T^{\mu (\rm eff)}_{\ \nu} \right] \,,
    \label{eq:Einstein_tensor}
\ee
where the effective energy-momentum tensor is defined as
\begin{align}
    T^{\mu (\rm eff)}_{\ \nu} =& \m^2 \left[ (1 - f) R^{\mu}_{\ \nu} + \nabla^\mu \nabla_\nu f + \frac{1}{2} \delta^{\mu}_{\ \nu} \left( f R - R - 2 \nabla_\sigma \nabla^\sigma f \right) \right.\nonumber \\
     & \left. \qquad + \delta_P \left( -\frac{3}{2} \frac{\nabla^\mu f \nabla_\nu f}{f} + \frac{3}{4} \delta^{\mu}_{\ \nu} \frac{\nabla_\sigma f \nabla^\sigma f}{f} \right)  \right] \,,
\end{align} 
Again, in the limit of general relativity $f(\phi) = 1$, this tensor vanishes, and the metric and Palatini formalisms reduce to the same field equations.

Next, we adopt the flat Friedmann--Lema\^itre--Robertson--Walker (FLRW) metric
\be 
\dd s^2 = g_{\mu\nu} \dd x^\mu \dd x^\nu = - \dd t^2 + a^2(t) \left[ \dd r^2 + r^2 \left( \dd \theta^2 + \sin^2\theta \dd \phi^2  \right)  \right] \,
\ee
where $a(t)$ is the cosmic scale and $t$ is the cosmic time.

From the $(0,0)$ and $(i,j)$ components of Eq.~\eqref{eq:Einstein_tensor} we obtain the modified Friedmann and Raychaudhuri equations
\begin{equation}
3 f H^2  = \frac{ \dot{\phi}^2/2 + V + \rho_m + \rho_r}{\m^2} - 3 H \dot{f} - \delta_{\rm P} \frac{3 \dot{f}^2}{4 f} \,,
\label{eq:Friedmann}
\end{equation}
\begin{equation}
-2 f \dot{H} = \frac{\dot{\phi}^2 + \rho_m + 4\rho_r/3}{\m^2} + \ddot{f} - H \dot{f} - \delta_{\rm P} \frac{3 \dot{f}^2}{2 f} \,,
\label{eq:Raychaudhuri}
\end{equation}
where $H = \dot{a}/a$ is the Hubble parameter, with a dot denoting the derivative with respect to cosmic time. 

The modified Klein-Gordon equation in either formulation reads 
\begin{equation} \label{eq:KGequation}
\ddot{\phi} + 3H \dot{\phi} + V_{,\phi} = \frac{\m^2}{2} f_{,\phi} \mathcal{R},
\end{equation}
where
\begin{equation}
\label{asdijbjfiawerwe}
\mathcal{R} = 6\left( \dot{H} + 2H^2 \right) + \delta_{\rm P} \left( - \frac{3 \dot{f}^2}{2 f^2} + 3 \frac{\ddot{f}}{f} + 9 H \frac{\dot{f}}{f} \right).
\end{equation}

Since matter and radiation are minimally coupled in~\eqref{eq:action}, their stress tensors are separately conserved~\cite{Koivisto:2005yk}
\begin{equation}
\dot\rho_m+3H\rho_m=0\,,\qquad
\dot\rho_r+4H\rho_r=0\,.
\label{eq:continuity}
\end{equation}

We introduce the fractional energy densities of radiation, matter, and the scalar field
\be 
    \Omega_r = \frac{\rho_r}{3 \m^2 f H^2} \,, \quad \Omega_m = \frac{\rho_m}{3 \m^2 f H^2}\,, \quad \Omega_\phi = \frac{\dot{\phi}^2}{6 f H^2} - \frac{f_{,\phi} \dot{\phi}}{f H} - \delta_P \frac{f_{,\phi}^2 \dot{\phi}^2}{4 f^2 H^2} + \frac{V(\phi)}{3 f H^2}
    \label{eq:energy_densities}
\ee
In the minimal coupling limit $f(\phi) = 1$, these expressions reduce to their standard GR definitions. With a non-minimal coupling, caution is required in interpreting these fractions, because the coupling mixes contributions in such a way that they are not guaranteed to remain $\leq 1$ (or even strictly positive). Under the standard assumptions that radiation and matter have non-negative energy densities ($\rho_r \geq 0\,, \ \rho_m \geq 0$) and $f(\phi) > 0$, $\Omega_r$ and $\Omega_m$ remain positive definite. For $\Omega_\phi$, the first three terms of Eq.~\eqref{eq:energy_densities} can be viewed as a relative kinetic contribution, while the last term represents the relative potential part. If the potential is non-negative, the potential contribution is always non-negative, whereas the kinetic contribution can become negative in certain non-minimal regimes.

It is useful to combine~\eqref{eq:Friedmann} and~\eqref{eq:Raychaudhuri} to obtain an expression for the Hubble flow parameter,
\be 
    \frac{\dot{H}}{H^2} = - \frac{1}{2 f H^2\m^2} \left( \dot{\phi}^2 + \rho_m +\frac{4}{3}\rho_r\right)  - \frac{\ddot{f}}{2 f H^2} + \frac{\dot{f}}{2 f H} + \delta_{\rm P} \frac{3}{4} \left( \frac{\dot{f}}{f H} \right)^2 \,.
\ee

For later comparison with $\Lambda$CDM and to keep the background equations in their standard GR form, it is convenient to define an effective dark sector with density and pressure
\begin{equation} \label{eq:effectivedensity}
    \rho_{\rm{eff}} = \frac{\rho_{r}}{f(\phi)} + \frac{\rho_{m}}{f(\phi)} + \rho_{\phi,{\rm eff}},
\end{equation}
\begin{equation}
    p_{\rm{eff}} = \frac{1}{3}\,\frac{\rho_{r}}{f(\phi)} + p_{\phi,{\rm eff}},
\end{equation}
where we have defined the effective density and pressure of the field as
\begin{equation}
    \label{asdfoubaiuewfwer}
    \rho_{\phi,{\rm eff}}=\frac{\m^{2}}{f(\phi)} \left[ \frac{\dot\phi^{2}}{2} + V(\phi) - 3H\,f_{,\phi}(\phi)\,\dot\phi - \delta_P \frac{3}{4} \frac{f_{,\phi}^2(\phi)}{f(\phi)} \dot\phi^{2} \right],
\end{equation}
and
\begin{equation}
    \label{asdofibaiuwer}
    p_{\phi,{\rm eff}} = \frac{\m^{2}}{f(\phi)} \left[ \frac{\dot\phi^{2}}{2} + 2H\,f_{,\phi}(\phi)\,\dot\phi + f_{,\phi\phi}(\phi)\,\dot\phi^{2} + f_{,\phi}(\phi)\,\ddot\phi +\delta_P \frac{3}{4} \frac{f_{,\phi}^2(\phi)}{f(\phi)} \dot\phi^{2} - V(\phi) \right].
\end{equation}
With these definitions the background equations take the GR form $3\m^2H^2=\rho_{\rm{eff}}$ and $3\m^2H^2+2\m^2\dot H=-p_{\rm{eff}}$, and the total equation of state, $w_{\rm tot}\equiv
p_{\rm{eff}}/\rho_{\rm{eff}}$, satisfies
\begin{equation}
w_{\rm tot} =-1-\frac{2}{3}\,\frac{\dot H}{H^2}= \frac{\Omega_r}{3} + \frac{1}{3fH^2}\left[ \frac{\dot\phi^{2}}{2} + 2Hf_{,\phi}\,\dot\phi + f_{,\phi\phi}\dot\phi^{2} + f_{,\phi}\ddot\phi +\delta_P \frac{3}{4} \frac{f_{,\phi}^2}{f} \dot\phi^{2} - V(\phi) \right]  \,.
\label{eq:weff}
\end{equation}
In the GR limit, the standard behaviors are recovered: during radiation domination $w_{\rm eff}=1/3$, during matter domination $w_{\rm eff}=0$, and under potential domination $w_{\rm eff}=-1$. Cosmic acceleration requires $w_{\rm eff}< -1/3$, while a phantom crossing $w_{\rm eff}< -1$ leads to $\dot H>0$. The de Sitter solution is characterized by $w_{\rm eff}=-1$ and a constant Hubble parameter $H=\sqrt{\Lambda/3}$. A minimally coupled scalar cannot sustain a phantom crossing since its expansion history is restricted to $-1\le w_{\rm eff}\le 1$, unless the Lagrangian contains a ghost term~\cite{Caldwell:1999ew, Carroll:2003st}. In contrast, a non-minimal coupling can support phases with $w_{\rm eff}<-1$~\cite{Gannouji:2006jm, Boisseau:2000pr, Torres:2002pe} without necessarily becoming unstable~\cite{Gunzig:2000kk, Faraoni:2001tq, Carvalho:2004ty, Carroll:2004hc, Perivolaropoulos:2005yv, Nesseris:2006er, Gannouji:2006jm}.

\subsection{Dynamical system analysis} \label{sec:dynamicalsystem}
Next, we recast the background equations into an autonomous system, a suitable language for both the numerical analysis and understanding the dynamics. We study the model by locating the critical points of this system, \textit{i.e.}~points where the derivatives of the phase-space variables vanish. To proceed, we introduce the standard $e$-fold time $N \equiv \ln a$ and the dimensionless parameters

\begin{figure}[h]
         \centering
         \includegraphics[width=0.7\textwidth]{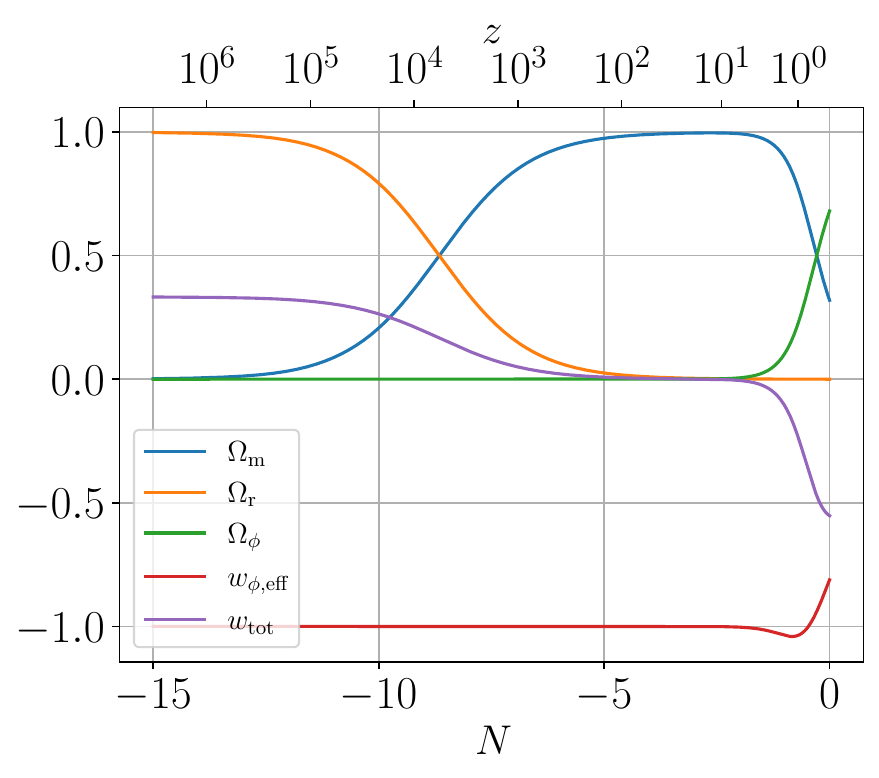}
     \caption{Time evolution of the density parameters of matter (blue), radiation (orange), and quintessence (green), as well as of the effective barotropic parameter of quintessence (red) and the Universe (purple), as a function of the elapsing number of e-folds $N$ and redshift $z$, for the CMB+DESI+DESY5 best-fit parameters $\xi=-1.41$, $\lambda = 1.95$, $\Omega_{\rm m}=0.3193$, $H_0=66.70$km$/$s$/$Mpc, and $\Omega_{\rm b}=0.05049$.}
     \label{fig:parameters}
\end{figure}
\be \label{eq:dynamicalsystemvariables}
x_1 = \frac{\dot{\phi}}{\sqrt{6}H\m}\,, \quad x_2 = \frac{\sqrt{V}}{\sqrt{3}H\m}\,, \quad x_3 = \frac{\rho_r}{3 H^2\m^2}\,, \quad x_4 = \frac{\phi}{\m} \,, \quad \lambda = - \m\frac{V_{,\phi}}{V}\,.  
\ee
and we keep the matter density fraction $\Omega_m \equiv \frac{\rho_m}{3 H^2\m^2}$ as an auxiliary variable. The Friedmann equation~\eqref{eq:Friedmann} gives a single algebraic constraint among $(x_1,x_2,x_3,x_4,\Omega_m)$: 
\be 
    \Omega_m = 1+ \xi x_4^2 - x_1^2 - x_2^2 - x_3  + 2 \sqrt{6} \xi x_1 x_4 + 6 \delta_{\rm P} \xi^2 \frac{x_1^2 x_4^2}{f} \,,
    \label{eq:constraint}
\ee
which defines a three-dimensional phase surface embedded in $\mathbb{R}^4$. Physical trajectories must also satisfy $f>0$.
Using the background equations \eqref{eq:Friedmann}–\eqref{eq:weff} one finds
the closed system for ${\bf X}=(x_1,x_2,x_3,x_4)$:
\bea 
    \frac{\dd x_1}{\dd N} &= \frac{\beta}{\sqrt{6}} + \epsilon x_1 \\
    \frac{\dd x_2}{\dd N} &= - \frac{\sqrt{6}}{2} \lambda x_1 x_2 + \epsilon x_2 \\
    \frac{\dd x_3}{\dd N} &= - 4 x_3 + 2 \epsilon x_3\\
    \frac{\dd x_4}{\dd N} &= \sqrt{6} x_1
    \label{eq:system}
\eea 
where 
\begin{align}
    &\beta\equiv \frac{\ddot{\phi}}{\m H^2}=-3\sqrt{6}x_1 + 3\lambda x_2^2 + \xi x_4\frac{\mathcal{R}}{H^2}=\nonumber\\
    &=\frac{1}{1+\frac{6\xi^2x_4^2}{f}(1-\delta_{\rm P})}\Bigg\{-3\sqrt{6}x_1 + 3\lambda x_2^2 +\xi x_4\Bigg[3-\frac{9}{f}(x_1^2-x_2^2)-\frac{9x_3}{f}+\nonumber\\
    &+\frac{36\xi x_1^2}{f}(\delta_{\rm P}-1)+\frac{6\sqrt{6}\xi x_4 x_1}{f}(3\delta_{\rm P}-2) + \frac{18\delta_{\rm P}\xi^2x_1^2x_4^2}{f^2}\Bigg]\Bigg\}
    \label{eq:beta}
\end{align}
and 
\begin{equation}
    \epsilon \equiv - \frac{\dot{H}}{H^2} =\frac{3}{2}+\frac{3}{2f}\left(x_1^2-x_2^2\right)+\frac{x_3}{2f}+\frac{6\xi x_1^2}{f}+\frac{2\sqrt{6}\xi x_1 x_4}{ f}-9\delta_{\rm P} \frac{\xi^2 x_1^2 x_4^2}{f^2} + \frac{\beta \xi x_4}{f}
    \label{eq:epsilon}
\end{equation}
From this, the effective barotropic parameter of the universe $w_{\rm tot}$ can be read off from the expression $\epsilon=3(1+w_{\rm tot})/2$ as 
\begin{equation}
    w_{\rm tot} = \frac{1}{A}\left(x_1^2-x_2^2\right)+\frac{x_3}{3f}+\frac{4\xi x_1^2}{f}+\frac{4\sqrt{6}\xi x_1 x_4}{3 f}-\delta_{\rm P} \frac{6\xi^2 x_1^2 x_4^2}{f^2} + \frac{2\beta \xi x_4}{3f}.
\end{equation}
The effective barotropic parameter of the field reads
\begin{equation}
    w_{\phi,{\rm eff}} = \frac{p_{\phi,{\rm eff}}}{\rho_{\phi,{\rm eff}}},
\end{equation}
where $p_{\phi,{\rm eff}}$ and $\rho_{\phi,{\rm eff}}$ are given by Eqs.~\eqref{asdofibaiuwer} and~\eqref{asdfoubaiuewfwer}, respectively.

In Fig.~\ref{fig:parameters} we present relevant cosmological parameters as a function of the number of e-folds and redshift, product of solving Eq.~\eqref{eq:system} numerically. The evolution is standard, following radiation-dominated (RD) and matter-dominated (MD) eras, until around $z\lesssim 10$, when quintessence begins to dominate. Its barotropic parameter becomes phantom for a brief period, before the present time when $w_{\phi,{\rm eff}}=-0.81$. 

Solving the fixed-point conditions 
\begin{equation}
    \frac{\dd x_1}{\dd N} = \frac{\dd x_2}{\dd N} = \frac{\dd x_3}{\dd N} = \frac{\dd x_4}{\dd N} = 0 \,,
\end{equation}
together with the algebraic constraint, yields three physically distinct branches: 
(i) A \emph{de Sitter} branch in which the scalar is frozen, matter and radiation vanish, and the expansion is characterized by $w_{\rm eff}=-1$ with $\Omega_\phi=1$. This branch splits into two algebraic solutions, denoted \textbf{DE-a} and \textbf{DE-b}. Both correspond to accelerated expansion; however, their existence conditions differ: \textbf{DE-a} exists whenever $\lambda^{2}\le 4\xi$ or for $\xi<0$ and remains physical with positive effective Planck mass (i.e.\ $f(\phi_*)>0$), whereas \textbf{DE-b} exists only if $\lambda^{2}\le 4\xi$ and becomes unphysical for $\xi<0$ because $f(\phi_*)<0$. 
(ii) A \emph{radiation} point \textbf{R}, with $\Omega_m=0$ and $w_{\rm eff}=1/3$. 
(iii) A \emph{matter} point \textbf{M}, with $\Omega_m=1$ and $w_{\rm eff}=0$. 
The explicit coordinates of these points in $(x_1,x_2,x_3,x_4)$, together with their existence conditions, are summarized in Table~\ref{tab:critpoints}.

\begin{table}[t]
\centering
\renewcommand{\arraystretch}{1.22}
\begin{tabular}{|c| c| l| c| c| c|}
\hline
C.P. & $(x_1^*,\;x_2^*,\;x_3^*,\;x_4^*)$ & Existence & $\Omega_\phi$ & $w_{\rm eff}$ & Acceleration\\
\hline\hline
DE-a &
$\Big(0,\ \dfrac{2}{\lambda}\sqrt{\,2\xi+\Delta\,},\ 0,\ \dfrac{-2\xi-\Delta}{\lambda\xi}\Big)$
& $\lambda^2\le 4\xi \ \text{or}\ \xi<0$ & $1$ & $-1$ & Yes \\[2mm]
DE-b &
$\Big(0,\ \dfrac{2}{\lambda}\sqrt{\,2\xi-\Delta\,},\ 0,\ \dfrac{-2\xi+\Delta}{\lambda\xi}\Big)$
& $\lambda^2\le 4\xi$ & $1$ & $-1$ & Yes \\[2mm]
M & $\big(0,\ 0,\ 0,\ 0\big)$ with $\Omega_m=1$ & always & $0$ & $0$ & No \\[1mm]
R & $\big(0,\ 0,\ 1,\ 0\big)$ & always & $0$ & $1/3$ & No \\
\hline
\end{tabular}
\caption{Critical points for our model. Here $\Delta=\sqrt{\xi(4\xi-\lambda^2)}$.}
\label{tab:critpoints}
\end{table}

The relation $\dd x_4 / \dd N = \sqrt6\,x_1$ forces $x_{1*}=0$ at any fixed point, so the \emph{kinetic} ($x_{1*} \neq 0, x_{2*}=0$) and \emph{scaling} ($x_{1*} \neq 0, x_{2*} \neq 0$) solutions present in the minimal models \cite{Copeland:1997et} are absent here. The only physical fixed points in a flat FLRW universe are DE-a (and possibly DE-b), M, and R. Also, in our model, late-time acceleration is generic. The non-minimal coupling generates de Sitter fixed points with $\epsilon_*=0$ (hence $w_{\rm eff}=-1$) independently of the steepness of the potential. This is in contrast to the minimal case, where late-time acceleration occurs only if $\lambda^2 < 2$~\cite{Copeland:1997et}.

To assess the fate of the fixed points, we linearize the autonomous system \eqref{eq:system} around each solution, $\mathbf X\to\mathbf X_*+\delta\mathbf X$, and study the first–order system $\delta\mathbf X' = J_*\,\delta\mathbf X$. Because all fixed points have $x_{1*}=0$, every term in $\beta$ and $\epsilon$ proportional to $x_1$ or $x_1^2$ vanishes at the point. As a result, the \emph{linear} spectra are identical in the metric and Palatini formulations. The eigenvalues of the Jacobian $J_*$, listed in Table~\ref{tab:eigs}, determine the nature of each point: a fixed point is (linearly) stable if the real parts are all negative, unstable if at least one is positive, and a saddle if they have mixed signs. For the two de~Sitter branches, we find one vanishing eigenvalue associated with a non-hyperbolic direction, but the remaining eigenvalues show that \textbf{DE-a} is stable on its center manifold and constitutes the unique late-time attractor of the system. The companion branch \textbf{DE-b} always possesses a positive eigenvalue and is therefore unstable. The standard matter point \textbf{M} is a saddle, while the radiation point \textbf{R} is unstable. Trajectories that originate in the radiation/matter region are repelled from \textbf{R} and \textbf{M} and flow towards the accelerating attractor \textbf{DE-a}, implying an asymptotic de~Sitter expansion.

\begin{table}[t]
\centering
\renewcommand{\arraystretch}{1.15}
\begin{tabular}{|c| l| c|}
\hline
Label & The eigenvalues $(\lambda_1,\lambda_2,\lambda_3,\lambda_4)$ & Stability \\
\hline\hline
DE-a &
$-4,\ 0,\ -\frac{3}{2}+\frac{3}{2}\sqrt{\,1-\frac{8}{3}\Delta\,},\ -\frac{3}{2}-\frac{3}{2}\sqrt{\,1-\frac{8}{3}\Delta\,}$ &
Stable \\
DE-b &
$-4,\ 0,\ -\frac{3}{2}+\frac{3}{2}\sqrt{\,1+\frac{8}{3}\Delta\,},\ -\frac{3}{2}-\frac{3}{2}\sqrt{\,1+\frac{8}{3}\Delta\,}$ &
Unstable \\
M &
$-1,\ \frac{3}{2},\ \frac{1}{4}\!\left(-3-\sqrt{\,9+48\,\xi\,}\right),\ \frac{1}{4}\!\left(-3+\sqrt{\,9+48\,\xi\,}\right)$ &
Unstable \\
R &
$+1,\ -2,\ -5,\ 0$ &
Unstable \\
\hline
\end{tabular}
\caption{Eigenvalues of the linearized system. Here $\Delta=\sqrt{\xi(4\xi-\lambda^2)}$.
}
\label{tab:eigs}
\end{table}

\begin{figure}[h]
     \centering
     \begin{subfigure}[b]{0.495\textwidth}
         \centering
         \includegraphics[width=\textwidth]{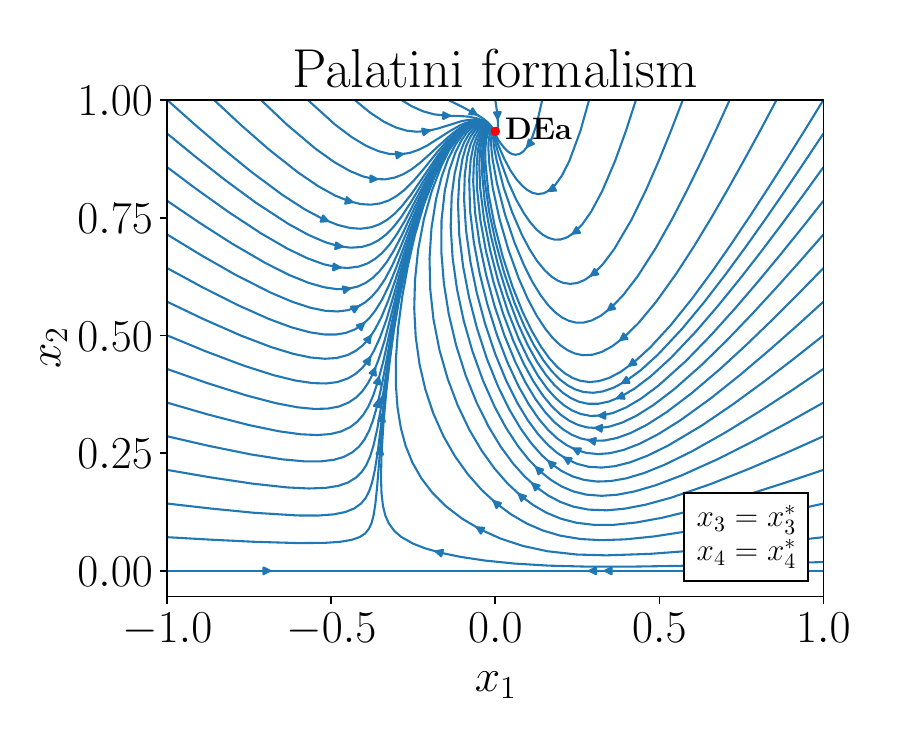}
     \end{subfigure}
     \begin{subfigure}[b]{0.495\textwidth}
         \centering
         \includegraphics[width=\textwidth]{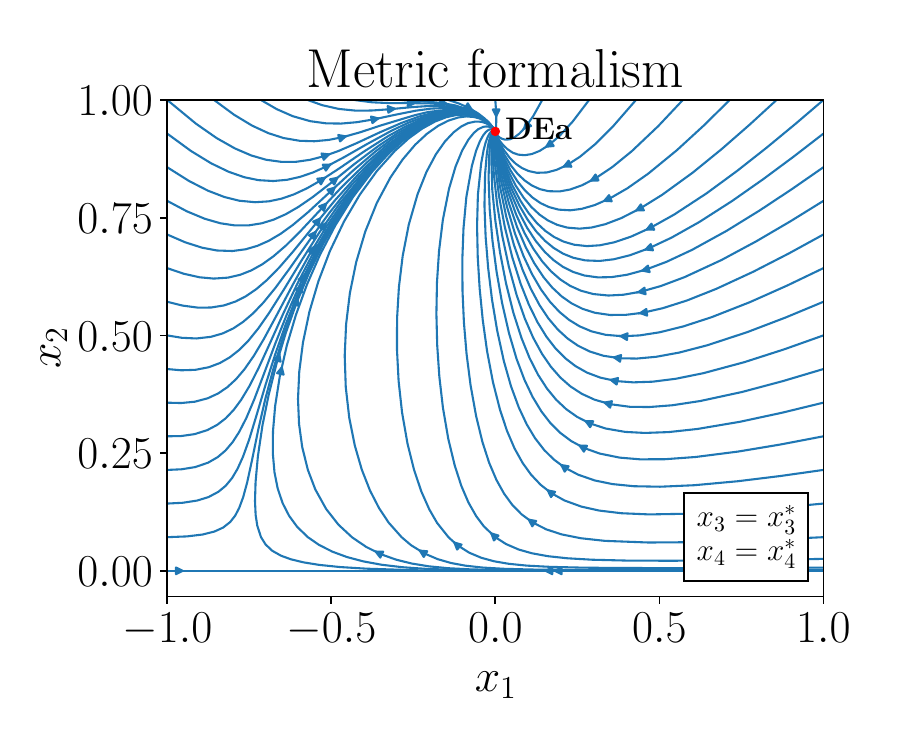}
     \end{subfigure}
     \caption{Phase space slice in the $(x_1,x_2)$ plane for the Palatini (left) and metric (right) formalisms, given the CMB+DESI+DESY5 best-fit parameters $\xi=-1.41$, $\lambda = 1.95$, $\Omega_{\rm m}=0.3193$, $H_0=66.70$km$/$s$/$Mpc, and $\Omega_{\rm b}=0.05049$. The variables $x_3$ and $x_4$ are fixed to their \textbf{DE-a} (red point) attractor values. Both formalisms share the same fixed points but the overall dynamics differ.}
     \label{fig:phasex1x2}
\end{figure}
\section{Datasets and Methodology}\label{asodfjbiawerwer}
In this section, we describe the data, as well as the methodology, used to constrain our model. We use three datasets: the second data release (DR2) of the baryon acoustic oscillations (BAO) distance measurements from the Dark Energy Spectroscopic Instrument (DESI)~\cite{DESI:2025zpo, DESI:2025zgx}, the cosmic microwave background (CMB) compressed likelihood~\cite{Zhai:2018vmm} (see also Ref.~\cite{Lemos:2023xhs}) obtained from the final Planck data release~\cite{Planck:2018vyg}, and type Ia supernovae (SNIa) distance moduli measurements from the Dark Energy Survey Year 5 (DESY5)~\cite{DES:2024jxu}.

\subsection{CMB Data}
The full CMB likelihood encodes information about dark energy perturbations mainly via the integrated Sachs-Wolfe (ISW) effect \cite{Sachs:1967er} and gravitational lensing. However, our model reduces to standard $\Lambda$CDM for $z\gtrsim 50$ (see below) and for $z< 50$, previous results~\cite{Fan:2016bgd} suggest that the effect of the non-minimal coupling term $\xi \phi^2$ on the ISW time integrals is small. It has also been argued~\cite{Wang:2025vfb} that not only the full CMB data impose weak constraints on the dark energy parameters, but may bias parameter estimation through prior sensitivity. We therefore employ the more robust CMB compressed data, in the same vein as the DESI collaboration~\cite{DESI:2025zgx} (see Appendix A therein), and many other works~\cite{Nesseris:2025lke, Pan:2025qwy, Berbig:2024aee, Akrami:2025zlb, Dinda:2025iaq, Giani:2025hhs, Hussain:2025nqy, Wang:2025vfb, Keeley:2025rlg, Gialamas:2025pwv, Li:2025ops, Dhawan:2025mer, Hogas:2025ahb, Mishra:2025goj, Braglia:2025gdo, Goldstein:2025epp, Gomez-Valent:2025mfl, Guedezounme:2025wav}. The CMB compressed data is comprised of the shift parameters $R$ and $l_{\rm a}$, together with $\omega_{\rm b} = \Omega_{\rm b}h^2$ ($h\equiv H_0/(100\, \text{km}\,\text{s}^{-1}\,\text{Mpc}^{-1}) $). These quantities are given by
\begin{equation}
    R = \sqrt{\Omega_{\rm m}H_0^2}\frac{D_M(z_{*})}{c}, 
\end{equation}
and
\begin{equation}
    l_{\rm a} = \pi \frac{D_M(z_{*})}{r_s(z_{*})}, 
\end{equation}
where $z_{*}$ is the redshift of photon decoupling and $\Omega_{\rm m}=\Omega_{\rm cmd} + \Omega_{\rm b}$, with $\Omega_{\rm cmd}$ and $\Omega_{\rm b}$ being the density parameters of cold dark matter and baryonic matter, respectively. To find $z_{*}$ we use the fitting formula from Ref.~\cite{Hu:1995en}, given by
\begin{equation}
    \label{eq:hufitting}
    z_{*} = 1048 \left(1 + 0.00124\omega_{\rm b}^{-0.738}\right)\left(1 + g_1\omega_{\rm m}^{g_2}\right),
\end{equation}
\begin{figure}[h]
     \centering
     \begin{subfigure}[b]{0.45\textwidth}
         \centering
         \includegraphics[width=\textwidth]{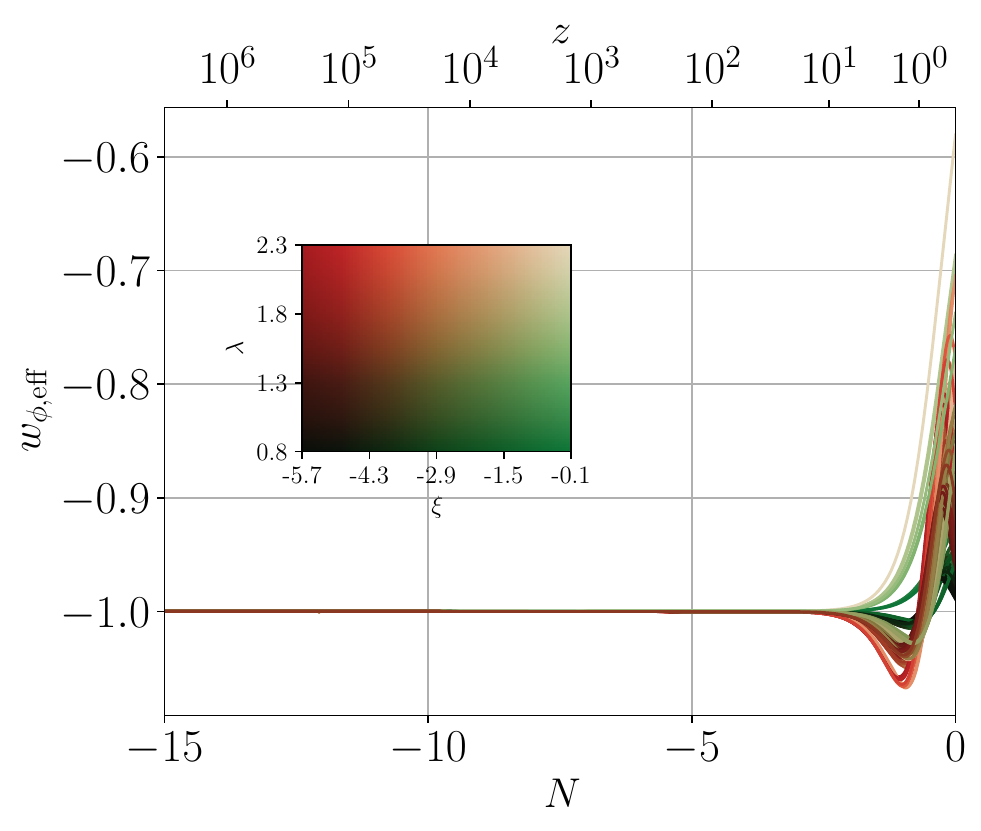}
     \end{subfigure}
     \begin{subfigure}[b]{0.45\textwidth}
         \centering
         \includegraphics[width=\textwidth]{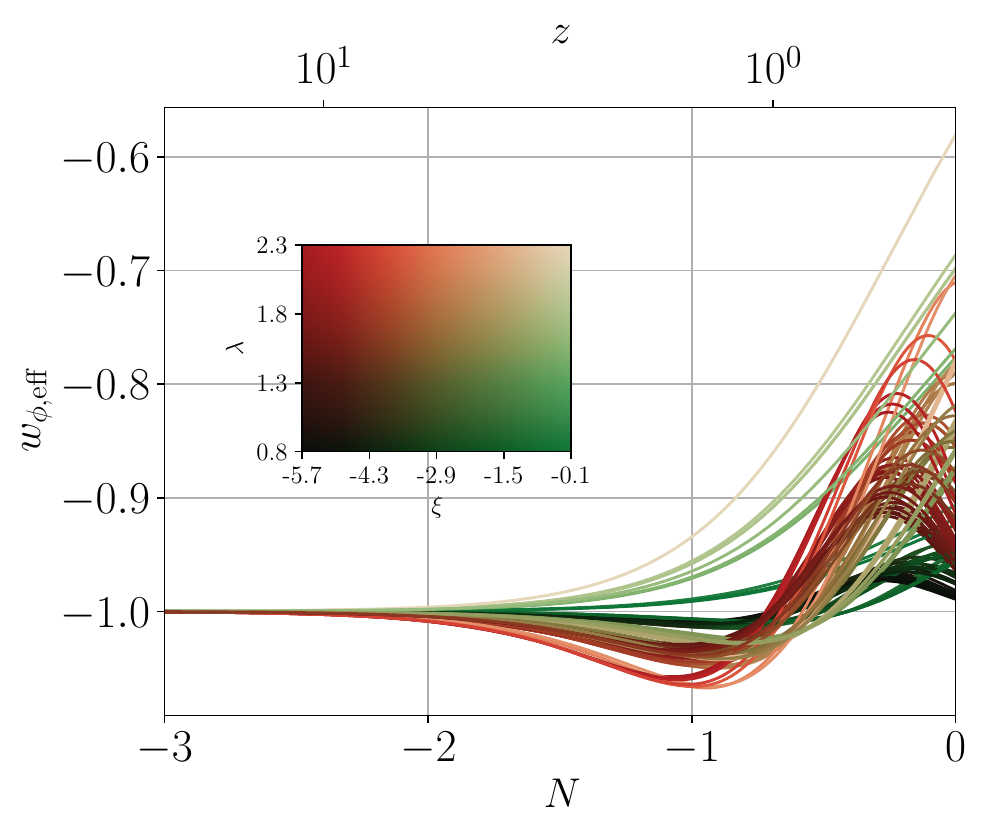}
     \end{subfigure}
          \begin{subfigure}[b]{0.45\textwidth}
         \centering
         \includegraphics[width=\textwidth]{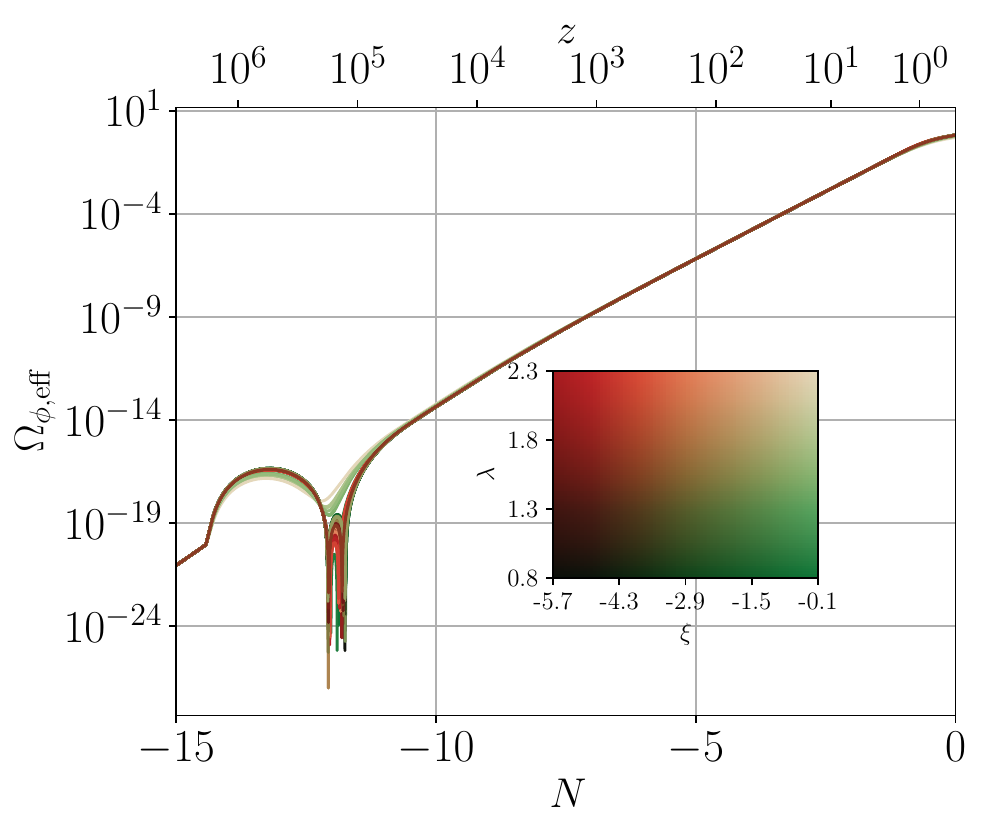}
     \end{subfigure}
     \begin{subfigure}[b]{0.45\textwidth}
         \centering
         \includegraphics[width=\textwidth]{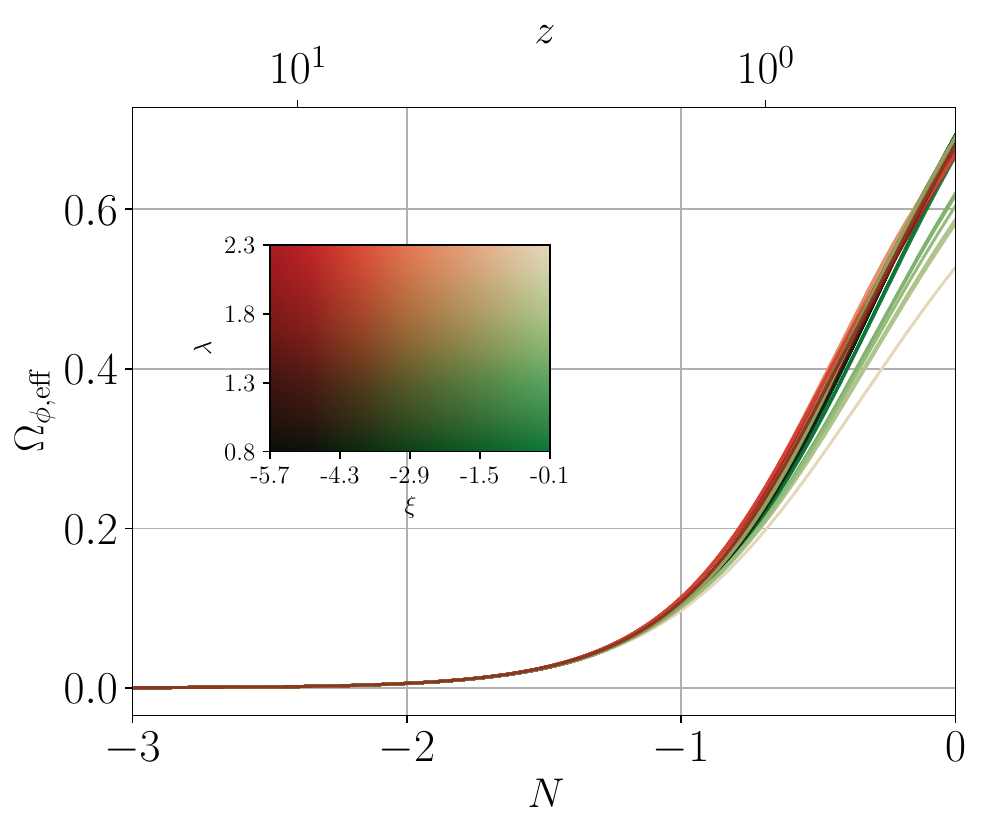}
     \end{subfigure}
     \caption{Upper Left: 100 realizations of the effective barotropic parameter of non-minimally coupled quintessence with an exponential potential for the entire integration range $N=[-15,0]$. The values of $\xi$ and $\lambda$ are randomly drawn from the joint 95\% credible region in $(\xi,\lambda)$. The rest of the model parameters are fixed to $x_2(-15)=3.00\times 10^{-11}$, $H_0=66.70$ and $\Omega_{\rm b}=0.05050$. Upper Right: Zoomed-in version of the upper left panel focusing on $N=[-3,0]$. Bottom left: 100 realizations of the effective density parameter for the same parameter values as the upper panels, for the entire integration range $N=[-15,0]$. Bottom right: Zoomed-in version of the lower left panel focusing on $N=[-3,0]$.}
     \label{fig:manywandome}
\end{figure}
where $g_1 = 0.0783\omega_{\rm b}^{-0.238}/\left(1 + 39.5\omega_{\rm b}^{0.763}\right)$, $g_2 = 0.560/\left(1 + 21.1\omega_{\rm b}^{1.81}\right)$, and $\omega_{\rm m}=\Omega_{\rm m}h^2$. Eq.~\eqref{eq:hufitting} depends only on pre-recombination physics. Its use is justified since in our model the effects of quintessence and, therefore, of modified gravity, only become relevant at low redshift. To demonstrate this, we plot $w_{\phi, {\rm eff}}$ and $\Omega_{\phi, {\rm eff}}$ for 100 random pairs of $\xi$ and $\lambda$ (taken from the joint $(\xi,\lambda)$ 95\% credible region) in Fig.~\ref{fig:manywandome}. It is apparent that for $z\gtrsim 50$ (which is much smaller than $z_{*}\sim 1100$), the effect of the field reduces to that of a subdominant cosmological constant. Furthermore, since the field is initially frozen at zero (see below), $f=1$, and there are no modifications to the effective Planck mass until late times. Even then, the total field excursion remains sub-Planckian $\Delta \phi/\m<1$, as can be seen from the bottom right panel of Fig.~\ref{fig:confidenceintervals}, where we plot the posterior distribution of $x_4=\phi/\m$ at present. Since the field is initially fixed at zero, indeed $x_4(z=0)=\Delta \phi/\m$.

$D_M(z)$ is the transverse comoving distance, given by (for a nearly-flat FLRW universe)
\begin{equation}
    D_M(z)=\frac{c}{H_0 \sqrt{\Omega_k}}\sinh{\left[ \sqrt{\Omega_k} \int_0^z \frac{\text{d}z'}{h(z')}\right]}\overset{\abs{\Omega_k} \ll 1}{\simeq} \frac{c}{H_0}\int_0^z \frac{\text{d}z'}{h(z')},
\end{equation}
where $h(z)\equiv H(z)/H_0$ is the reduced Hubble parameter. The comoving sound horizon, $r_s(z)$, is given by
\begin{equation}
    r_s(z)= \frac{1}{H_0}\int_{z}^{\infty}\frac{c_s(z')\text{d}z'}{h(z')}= \frac{c}{H_0}\int_{z}^{\infty}\frac{\text{d}z'}{\sqrt{3\left(1+\frac{3\Omega_{\rm b}}{4\Omega_{\rm \gamma}(1+z')}\right)}h(z')},
\end{equation}
where $\Omega_{\rm \gamma}$ is the density parameter of radiation, fixed by the temperature of the CMB via $\Omega_{\gamma}h^2 = (2.4729 \pm 0.0002)\times 10^{-5}$ \cite{ParticleDataGroup:2022pth}. Note that current bounds on the neutrino masses imply they become non-relativistic after decoupling, and so their contribution to the radiation density needs to be included in $h(z)$ when calculating the comoving sound horizon.

We employ a Gaussian prior on $x = (R,l_{\rm a}, \omega_{\rm b})$, with mean values given by
\begin{equation}
    \bar{x} = (1.74963, 301.80845, 0.02237),
\end{equation}
and the covariance matrix given by
\begin{equation}
    \mathcal{C} = 10^{-8}\times 
    \begin{bmatrix}
    1598.9554 & 17112.007 & - 36.311179 \\
    17112.007 & 811208.45 & - 494.79813 \\
    -36.311179 & - 494.79813 &  2.1242182
    \end{bmatrix}.
\end{equation}

\subsection{BAO Data}
The BAO sample provided by DESI in their second data release\footnote{The data and covariance matrix can be found in the following public repository \url{https://github.com/CobayaSampler/bao_data/tree/master/desi_bao_dr2}.} includes measurements of galaxies, quasars, and Lyman-$\alpha$ forest, spanning redshifts in the range $0.295\leq z \leq 2.33 $. They determine the ratios $D_M(z)/r_{\rm s}(z_{\rm d})$, $D_H(z)/r_{\rm s}(z_{\rm d})$, and $D_V(z)/r_{\rm s}(z_{\rm d})$, where 
$D_H(z)$ is the Hubble distance, given by
\begin{equation}
    D_H(z)=\frac{c}{H_0 h(z)},
\end{equation}
$D_V(z)$ is the isotropic BAO distance, given by 
\begin{equation}
    \label{eq:dvrs}
    D_V(z)=\left(z D_M^2(z)D_H(z)\right)^{1/3},
\end{equation}
and $z_{\rm d}$ is the redshift at the drag epoch. In order to be consistent with Ref.~\cite{DESI:2025zpo}, we use the following expression for the comoving sound horizon at the drag epoch~\cite{Brieden:2022heh}
\begin{equation} \label{adifbauywefwe}
    r_{\rm s}(z_{\rm d}) = 147.05 \, \text{Mpc}\left(\frac{\omega_{\rm b}}{0.02236}\right)^{-0.13}\left(\frac{\omega_{\rm m}}{0.1432}\right)^{-0.23}\left(\frac{N_{\rm eff}}{3.04}\right)^{-0.1},
\end{equation}
where $N_{\rm eff}$ is the effective number of relativistic neutrino species. Importantly, this equation assumes standard pre-recombination physics, making its use consistent with the fact that the field reduces to an effective cosmological constant for $z\gtrsim 50$ (see Fig.~\ref{fig:manywandome}). In our numerical analysis, we utilize the value $N_{\rm eff} = 3.044$ from recent neutrino decoupling simulations~\cite{Akita:2020szl, Froustey:2020mcq, Bennett:2020zkv, Drewes:2024wbw}.

\subsection{SNIa Data}
The SNIa sample provided by DESY5\footnote{The data, the covariance matrix, and the likelihood, can be found in the following public repository \url{https://github.com/des-science/DES-SN5YR}.} includes measurements of the distance modulus $\mu(z)$ for 1635 photometrically-classified supernovae in the redshift range $0.10 \leq z \leq 1.13$, as well as 194 low-redshift supernovae in the redshift range $0.024 \leq z \leq 0.10$. The distance modulus is given by
\begin{equation}
    \label{iajbjfdghiabdf}
    \mu(z) = 5\log_{10}\left[\frac{D_L(z)}{\text{Mpc}}\right] + 25,
\end{equation}
where $D_L(z)$ is the luminosity distance, related to the transverse comoving distance via $D_L(z) = (1+z) D_M(z)$. Since the absolute magnitude $M$ of SNIa is fully degenerate with $H_0$ (appearing in $D_L(z)$ through $D_M(z)$), both parameters can be combined as $\mathcal{M} = M + 5\log_{10}(c/H_0/\text{Mpc})$. In the computation of the likelihood, the nuisance parameter $\mathcal{M}$ is marginalized over (see Appendix A.1 of Ref. \cite{Goliath:2001af} for further details).

\subsection{Dynamical System and Initial Conditions}
From the above, it is clear that the dynamical properties of our model are constrained by the data only via the function $h(z)$. To obtain it, we solve the dynamical system in Eq.~\eqref{eq:system} numerically. We begin the integration deep in the RD era at $N=-15$, \textit{i.e.}, 15 e-folds before the present time, at $N=0$. Along with Eq.~\eqref{eq:system}, we simultaneously solve Eq.~\eqref{eq:epsilon}, which may be recast as
\begin{equation}
    \frac{\text{d}\log H}{\text{d}N} = -\epsilon(x_1,x_2,x_3,x_4),
\end{equation}
where $\epsilon(x_1,x_2,x_3,x_4)$ summarizes the complicated expression in the right-hand-side. This equation benefits from a re-scaling invariance given by $H(N) \rightarrow c H(N)$. Using it, for any initial condition $H_{\rm ini}=H(-15)$, one may always normalize the solution by its value at present, $c = 1/H(0)$, to trivially obtain $h(N)$. Since Eq.~\eqref{eq:epsilon} is simultaneously solved with Eq.~\eqref{eq:system}, the obtained $h(z)$ automatically inherits the non-trivial late-time dynamics of non-minimally coupled quintessence. 

We make sure the density parameter of radiation is fixed to $\Omega_{\gamma}h^2 = x_3(0)h^2 = 2.4729 \times 10^{-5}$ by using a shooting algorithm. This effectively fixes the initial condition for the variable $x_3$. Furthermore, in the RD era, the energy densities satisfy $\rho_{\rm r} \gg \rho_{\phi,{\rm eff}}, \rho_{\rm m}$. From the Friedmann equation \eqref{eq:Friedmann} together with Eq.~\eqref{eq:effectivedensity}, this implies that $x_3 \simeq 1 \gg x_1,x_2,x_4,\Omega_{\rm m}$, which in turn implies that $\beta \ll 1$. Therefore, expressing the Ricci scalar in terms of the dynamical system variables, we find
\begin{align} \label{eq:ricciscalar}
    \frac{\mathcal{R}}{H^2}&= 3 (1-3w_{\rm eff}) + \frac{\delta_P}{A}\left(18\sqrt{6}\xi x_1 x_4 + 36\xi x_1^2 + 6\xi \beta x_4 - \frac{36\xi^2 x_1^2 x_4^2}{A}\right)= \nonumber \\
    &=\frac{3}{A}\Bigg[1 + x_4^2 - 3(x_1^2 - x_2^2) - x_3 + 12(\delta_P - 1)\xi x_1^2 +6\sqrt{6}\left(\delta_P - \frac{2}{3}\right)\xi x_1 x_4 +\nonumber\\
    & +2(\delta_P -1)\xi\beta x_4 + \delta_P \frac{6\xi^2 x_1^2 x_4^2}{A}\Bigg]\ll 1.
\end{align}
Furthermore, since $V\ll \m^2 H^2$, $V_{,\phi}=\frac{\lambda}{\m}V$ and $\lambda \sim \mathcal{O}(1)$, we have $V_{,\phi} \ll \m H^2$ and so the quintessence equation of motion \eqref{eq:KGequation} is reduced to
\begin{equation}
    \phi'' + (3 - \epsilon)\phi'=0,
\end{equation}
where primes denote derivatives with respect to the number of e-folds. Using that during RD $\epsilon = 2$, the field evolution is found to be
\begin{equation}
    \phi(N) = c_1 e^{-N} + c_2,
\end{equation}
where $c_1$ and $c_2$ are integration constants. We conclude that the field quickly freezes to a constant value, indeed corresponding to the thawing quintessence regime. We may set the initial condition $x_1(-15)=0$ without loss of generality. 
\begin{table}[H]
    \centering
    \begin{tabular}{|l|c|c|}
        \hline
        Parameter & Description & Prior \\
        \hline\hline
        $\xi$ & Non-minimal coupling & $\mathcal{U}$[-10, 0] \\
        $\lambda$ & Slope of the exponential potential & $\mathcal{U}$[0, 5] \\
        $x_2(-15)$ & Initial condition for $x_2$ & $\mathcal{U}$[$10^{-15}$, $10^{-10}$] \\
        $H_0$ & Hubble constant [km s$^{-1}$ Mpc$^{-1}$] & $\mathcal{U}$[60,80] \\
        $\Omega_{\rm b}$ & Density parameter of baryonic matter & $\mathcal{U}$[0, 0.1] \\
        \hline
    \end{tabular}
        \caption{Priors adopted for the cosmological parameters in the MCMC analysis.}
        \label{table:priors}
\end{table}
As for the initial condition on $x_2$, we leave $x_2(-15)$ as a free parameter. Of course, it should be small enough so that the system starts near the RD fixed point. This is ensured by imposing appropriate priors. The value of $x_2(-15)$ is directly related to $\Omega_{\rm m}$ (see below).

This leaves $x_4$. Although in order to begin the integration near the RD fixed point one needs $x_4\ll 1$, its value is not necessarily zero. However, we may note that in the minimally coupled case, $x_4$ is degenerate with $V_0$ (see Eq.~\eqref{eq:potential}), which means that one may set $x_4(-15)=0$ and vary $x_2(-15)$ to explore the totality of parameter space. Turning on the non-minimal coupling $\xi \neq 0$ breaks this degeneracy. Nevertheless, since $x_4\ll 1$ is still a requirement, the sensitivity of the dynamics to its precise initial value is expected to be suppressed. We thus set $x_4(-15)=0$ and leave an examination of the dependence on this initial condition to future work.

\subsection{Free Parameters and Posterior Sampling}
The parameter space of $\xi\phi$CDM is five-dimensional, $\Theta = \{\xi, \lambda, x_2(-15), H_0, \Omega_{\rm b}\}$. Those of $\phi$CDM and $\Lambda$CDM are nested subsets of $\Theta$, obtained by taking $\xi = 0$ and $\xi = \lambda = 0$, respectively. To obtain the posterior distribution of the parameters of each model, we perform a Markov Chain Monte Carlo (MCMC) analysis by using the publicly available \texttt{Python} package \texttt{emcee}~\cite{Foreman-Mackey:2012any}. The combined likelihood $\mathcal{L}$ from all three observational datasets used in this work is given by
\begin{equation} \label{eq:likelihood}
    -2 \log{\mathcal{L}}(\Theta) \equiv \chi_{\rm TOT}^2(\Theta) = \chi_{\rm CMB}^2(\Theta) + \chi_{\rm BAO}^2(\Theta) + \chi_{\rm SN}^2(\theta).
\end{equation}
For CMB and BAO, we use multivariate Gaussian likelihoods,
\begin{equation}
    \chi^2(\Theta) = \left(d_i - t_i(\Theta)\right)^T\mathcal{C}^{-1}_{ij}\left(d_j - t_j(\Theta)\right),
\end{equation}
where $d_i$ is the data vector, $t_i$ is the theoretical prediction vector for a given $\Theta$, and $\mathcal{C}_{ij}$ is the covariance matrix of each experiment. For the SNe likelihood, we follow DESY5, including a marginalization over $\mathcal{M}$ among other corrections. From Eq.~\eqref{iajbjfdghiabdf} follows that SNIa cannot constrain $\Omega_{\rm b}$ (which mainly enters $\chi^2_{\rm TOT}$ via the comoving sound horizon) or $H_0$ (which is degenerate with $M$). Therefore, the SN likelihood depends only on the subset $\theta = \{\xi,\lambda, x_2(-15)\}$. We impose uniform priors for all parameters, listed in Table~\ref{table:priors}.

The initial state of the chains is given by 128 walkers, following a Gaussian distribution centered on a preliminary best fit point, with a spread large enough to probe the entire prior space. Each walker is evolved for $8\times 10^{4}$ iterations in $\phi$CDM and $\Lambda$CDM,  and for $10^5$ in $\xi\phi$CDM. We assess convergence by following Refs. \cite{Foreman-Mackey:2012any,Goodman:2010dyf} to estimate the integrated autocorrelation time $\tau_{\rm p}$ for each parameter. After discarding a burn-in of $5\tau$, where $\tau=\text{max}(\tau_{\rm p})$, the number of remaining samples per walker $N$ must satisfy $N>50\tau$. 
 \begin{table}[H]
    \centering
    \small
    \begin{tabular}{|>{\centering\arraybackslash}p{1.0cm}|>{\centering\arraybackslash}p{4.1cm}|>{\centering\arraybackslash}p{4.1cm}|>{\centering\arraybackslash}p{4.1cm}|}
        \hline
         Params & $\Lambda$CDM & $\phi$CDM                                & $\xi\phi$CDM \\
        \hline\hline
        $\xi$       & --                               & --                     & $-2.50(-1.41)^{+1.7}_{-0.42}$\\
        
        $\lambda$   & --                               & $0.81(0.85)^{+0.18}_{-0.11}$ & $1.68(1.95)^{+0.36}_{-0.41}$\\
        
        $\Omega_{\rm m}$ & $0.3039(0.3039)\pm0.0038$         & $0.3153(0.3160)\pm0.0056$    & $0.3179(0.3193)\pm0.0059$\\
        
        $H_0$            & $68.43(68.43)\pm0.30$            & $67.00(66.91)\pm0.59$       & $66.82(66.70)\pm0.58$\\
        
        $\Omega_{\rm b}$ & $0.04791(0.04791)\pm0.00033$       & $0.05016(0.05029)\pm0.00090$  & $0.05036(0.05049)\pm0.00086$\\
        \hline
    \end{tabular}
        \caption{CMB+BAO+SN 68$\%$ ($1\sigma$) credible intervals and best-fit values (in parentheses) for the parameters of $\Lambda$CDM, $\phi$CDM and $\xi\phi$CDM. $H_0$ is given in units of $\text{km}/\text{s}/\text{Mpc}$.}
        \label{table:parameterconfidence}
\end{table}
Having ensured convergence, we take the initial chains and conservatively discard an initial 30$\%$ burn-in length (exceeding $5\tau$ in all models) to ensure the sampling is independent of the initial state. From the chains, we determine the matter density at present as a derived parameter. Posterior distributions are then plotted using the publicly available package \texttt{GetDist}~\cite{Lewis:2019xzd}. The maximum a posteriori (MAP) parameters (which coincide with their best-fit values since we assume flat priors) for each model are found by using a Nelder-Mead simplex minimizer~\cite{Nelder:1965zz} starting from the maximum likelihood sample obtained from MCMC.
\begin{table}[H]
    \centering
    \begin{tabular}{|l|l|c|c|c|c|}
        \hline
         & Model & Planck & DESI DR2 & DESY5 & Total \\
        \hline\hline
       \multirow{2}{*}{$\Delta \chi^2$} &$\xi\phi$CDM & 0.46 & -3.41 & -11.70 & -14.66\\
        &$\phi$CDM & 2.00 & -1.18 & -10.61 & -9.80\\
        \hline
        \multirow{2}{*}{$\log B$} &$\xi\phi$CDM & - & - & - & 5.52\\
        &$\phi$CDM & - & - & - & 2.45\\
        \hline
    \end{tabular}
        \caption{Change in $\chi^2$ (first two rows) and $\log B$ (last two rows) for $\xi\phi$CDM and $\phi$CDM relative to $\Lambda$CDM, evaluated at the CMB+BAO+SN best-fit parameters. The contribution from each dataset is shown in the corresponding column, with the total reported in the last column. A negative (positive) $\Delta \chi^2$ corresponds to an improvement (worsening) in fit. A positive (negative) $ \log B$ implies evidence in favor (against) of the model over $\Lambda$CDM.}
        \label{table:statisticssummary}
\end{table}
We assess the evidence for a given model $M$, defined as
\begin{equation}
    \mathcal{Z} = \int \text{d}\Theta \mathcal{L}(d|\Theta,M)p(\Theta|M),
\end{equation}
where the likelihood $\mathcal{L}(d|\Theta,M)$ is defined in Eq. \eqref{eq:likelihood} and $p(\Theta|M)$ is the prior, by using the publicly available package \texttt{MCEvidence}~\cite{Heavens:2017afc}. We do so by taking into account prior volumes and thinning the chains by $\tau/2$ to reduce autocorrelation between the samples that are used for the calculation. We interpret results according to the Jeffrey's scale~\cite{Jeffreys:1939xee, Trotta:2008qt}: given two models $M_0$ and $M_1$, with evidences $\mathcal{Z}_0$ and $\mathcal{Z}_1$, respectively, the evidence of $M_1$ over $M_0$ is determined inconclusive if $\log B_{M_1, M_0}<1.0$, where $B_{M_1, M_0}\equiv\mathcal{Z}_1 /\mathcal{Z}_0$ is the Bayes factor. It is weak for $1.0 < \log B_{M_1, M_0}< 2.5$, moderate for $2.5 < \log B_{M_1, M_0}< 5.0$ and strong for $\log B_{M_1, M_0}> 5.0$.
\begin{figure}[h]
         \centering
         \includegraphics[width=1\textwidth]{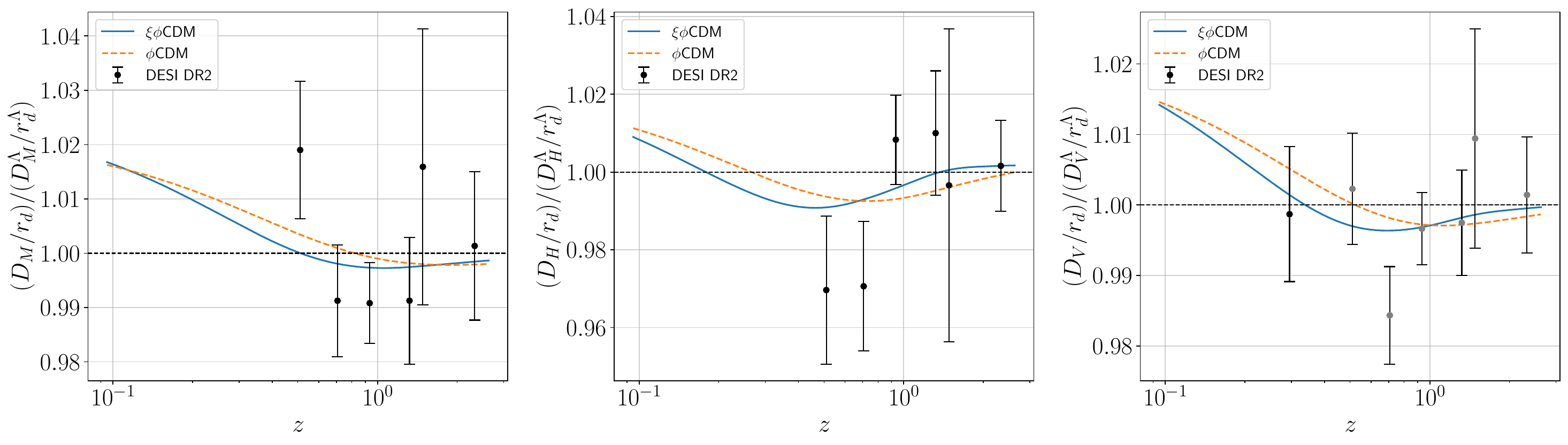}
     \caption{BAO distances $D_M(z)/z_d$ (left) $D_H(z)/z_d$ (center) and $D_V(z)/z_d$ (right) predicted by $\xi\phi$CDM (full blue) and $\phi$CDM (dashed orange) relative to $\Lambda$CDM (horizontal dashed black). We use the CMB+BAO+SN best-fit parameter values for each model. Black dots represent the DESI DR2 residuals. In gray are $D_V(z)/z_d$ values derived from the $D_M(z)/z_d$ and $D_H(z)/z_d$ actual data points.}
     \label{fig:desiresiduals}
\end{figure}

\section{Results} \label{sec:results}
In this section, we compare the predictions of our models with the data and analyze the constraints on the parameters obtained from sampling their posterior distributions. 

\subsection{Observational constraints} \label{easdfbahuisdfawew}
In Table~\ref{table:parameterconfidence} we report the CMB+BAO+SN 68$\%$ confidence intervals for the parameters of $\xi\phi$CDM, $\phi$CDM, and $\Lambda$CDM, along with their best-fit values. The $\phi$CDM best-fit value of $\lambda$ satisfies $\lambda<\sqrt{2}$. This agrees with the late-time dominant attractor condition~\cite{Copeland:1997et}, leading to an accelerating phase of expansion of the Universe. We emphasize that, in this attractor, minimally coupled quintessence with an exponential potential with $\lambda>\sqrt{2}$ cannot support acceleration. As our phase-space analysis reveals, this condition is relaxed with the addition of a non-minimal coupling. Indeed, the existence of the late-time acceleration attractor is guaranteed as long as $\xi<0$, independently of the value $\lambda$ takes. This is consistent with the reported $\lambda = 1.95$ best-fit value for $\xi\phi$CDM.

In Table~\ref{table:statisticssummary} we report the goodness of fit of $\xi\phi$CDM and $\phi$CDM relative to $\Lambda$CDM in terms of the difference in $\chi^2$, for each dataset as well as the total, given the best-fit parameters. We also provide the (logarithm of the) Bayes factor for both models. While we find that the $\Lambda$CDM baseline model is marginally in better agreement with the CMB data, the dynamical dark energy models discussed in our analysis significantly improve the fit to the BAO and SN data compared to the baseline in a joint analysis. We find the improvement in $\chi^2$ to be $\Delta \chi^2 = -14.66$ for $\xi\phi$CDM and $\Delta \chi^2 = -9.8$ for $\phi$CDM. Regarding $\log B$, the data shows weak (albeit close to moderate) evidence for $\phi$CDM over $\Lambda$CDM and strong evidence for $\xi\phi$CDM over $\Lambda$CDM. We investigate these claims further by analyzing the predictions of the models.

In Fig.~\ref{fig:desiresiduals} we plot the three BAO distances $D_M(z)/z_d$, $D_H(z)/z_d$, and $D_V(z)/z_d$ predicted by $\xi\phi$CDM and $\phi$CDM relative to $\Lambda$CDM. For all three models, we take their corresponding CMB+BAO+SN best-fit parameter values. We also show the DESI DR2 data relative to the best-fit $\Lambda$CDM model. Since we plot ratios of distances, deviations from $\Lambda$CDM correspond to deviations from unity. Note that DESI DR2 provides one data point only for $D_V(z)/z_d$, at $z=0.295$. The other six points in the third panel are obtained by using Eq.~\eqref{eq:dvrs} with the data for $D_M(z)/z_d$ and $D_H(z)/z_d$ at each redshift bin, with the corresponding propagated errors. These points are represented in gray to emphasize the difference with actual data. From the figure, it is clear that the largest improvement in fit is driven by the first two $D_H(z)/z_d$ data points, at redshifts $z=0.510$ and $z=0.706$, respectively.
\begin{figure}[h]
     \centering
        \includegraphics[width=0.7\textwidth]{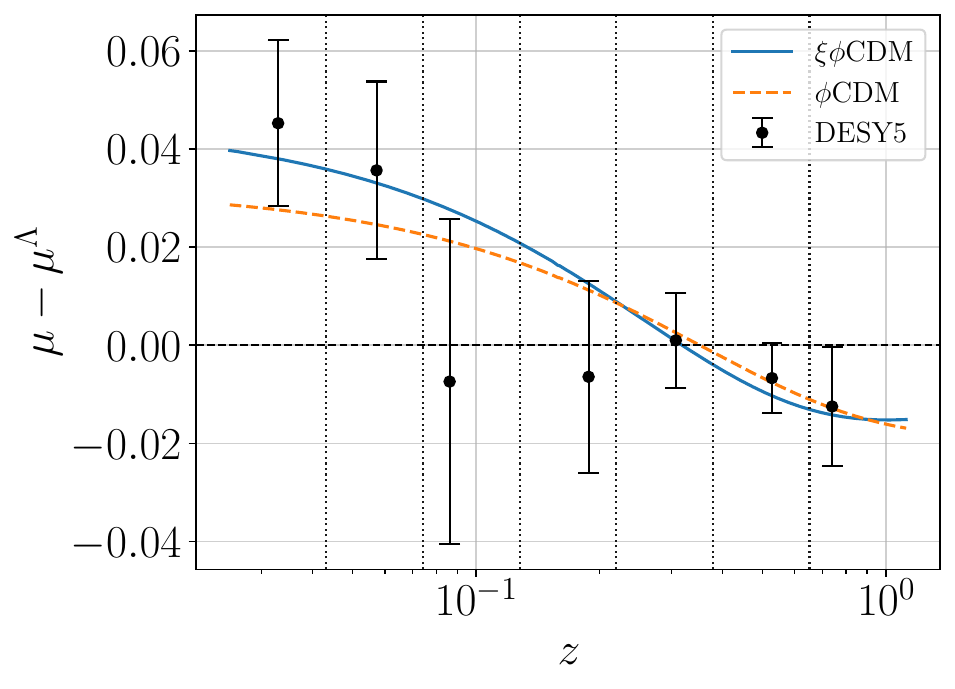}
     \caption{Distance moduli $\mu(z)$ predicted by $\xi\phi$CDM (blue) and $\phi$CDM (orange) relative to $\Lambda$CDM (horizontal dashed black). We use the CMB+BAO+SN best-fit parameter values for each model. 39 out of the total 1829 redshift values in the DESY5 data are duplicated. To construct the blue and orange curves, a small shift $\Delta z=5\times 10^{-8}$ was added to those points to allow interpolation. Black dots represent the DESY5 binned residuals and vertical dotted lines mark bin edges. The DESY5 data have been calibrated by subtracting the inverse-covariance weighted mean.}
     \label{fig:desyresiduals}
\end{figure}
In Fig.~\ref{fig:desyresiduals} we show the distance moduli predicted by $\xi\phi$CDM and $\phi$CDM, as well as the DESY5 data, all relative to $\Lambda$CDM. Again, for each model, we use their respective CMB+BAO+SN best-fit parameter values. For visualization purposes, we show the SN data binned in redshift, following a similar procedure to Ref.~\cite{DESI:2025zpo}. More specifically, given a residual vector $r_i=\mu_i^{\rm data}-\mu_i^{\Lambda\text{CDM}}$, where $i$ ranges from 1 to the total number $N$ of data points, a bin is a collection of indices ordered in ascending redshift
\begin{equation}
    b = \{j,...,j+k\},\quad \text{with}\quad  j,k>0\quad \text{and} \quad j+k<N.
\end{equation}
Given indices $p,q$ in the bin $p,q\in b$, we define $\hat{u}_p=1$ and $\hat{C}_{pq}$ as the submatrix in the bin of the total covariance matrix $C_{ij}$. With this, the residuals shown in the figure are given by
\begin{equation}
    \hat{r}_{b} = \frac{\hat{u}_{p}\hat{C}^{-1}_{p q}r_{q}}{\hat{u}_{p}\hat{C}^{-1}_{pq}\hat{u}_{q}},\quad \hat{\sigma}_{b} = \frac{1}{\sqrt{\hat{u}_{p}\hat{C}^{-1}_{pq}\hat{u}_{q}}}.
\end{equation}
We also account for the uncalibrated absolute magnitude $M$ of the SNe. To do so, we remove the weighted average from the data using the full covariance matrix. For each model, we also remove its corresponding offset relative to $\Lambda$CDM. This procedure ensures that all residuals (both data and models) with respect to $\Lambda$CDM have null weighted means
\begin{equation}
    u_i C_{ij}^{-1} r_j = 0, \quad \text{where} \quad u_i=1 \quad \text{and}\quad i,j=1, ..., N,
\end{equation}
so that they all share the same zero point. Finally, we emphasize that binning ignores cross-bin correlations and mixes within-bin correlations, making it useful for visualization purposes only. Keeping this caveat in mind, the figure suggests that $\xi\phi$CDM and $\phi$CDM are able to fit both the low- and high-redshift SNe better than $\Lambda$CDM. 
\begin{figure}[h]
         \centering
         \includegraphics[width=1\textwidth]{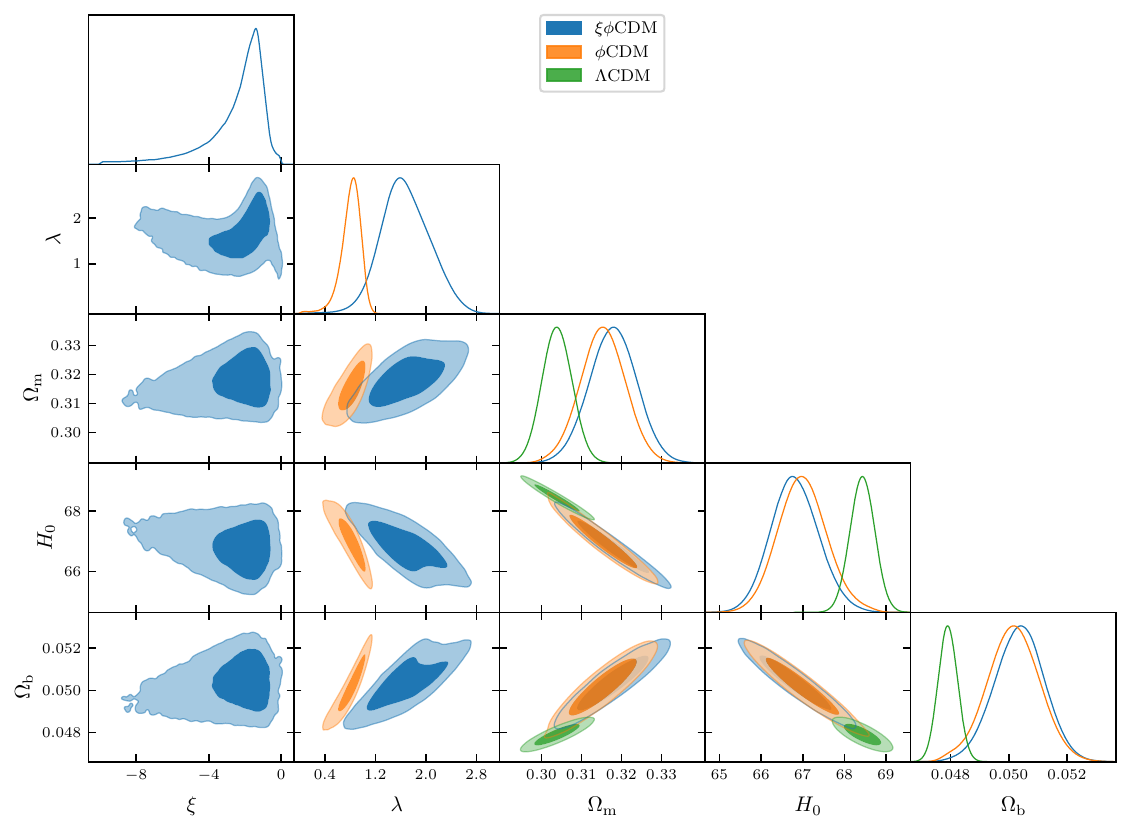}
     \caption{Parameter posteriors for $\xi\phi$CDM (blue), $\phi$CDM (orange), and $\Lambda$CDM (green) using CMB+DESI+DESY5. The darker (lighter) shaded regions represent the 68$\%$ (95$\%$) credible intervals.}
     \label{fig:cornerpalatini}
\end{figure}
Fig.~\ref{fig:cornerpalatini} shows the posterior distributions for the parameters of all three models. Indeed, since $\phi$CDM corresponds to $\xi=0$ and $\Lambda$CDM to $\xi=\lambda=0$ they can share the same figure. All posterior distributions lie within the prior range, indicating that our choice of priors does not drive the results. It is clear that the data prefers a non-zero value for the non-minimal coupling. This is one of the main results of our paper. Furthermore, at the start of this section we commented on the $\xi\phi$CDM and $\phi$CDM best-fit values of $\lambda$ satisfying their respective late-time acceleration attractor conditions. The same can be found in the $\xi\phi$CDM posterior distributions, where most of the 2$\sigma$ confidence interval of $\lambda$ lies at $\lambda>\sqrt{2}$.
\begin{figure}[h]
     \centering
     \begin{subfigure}[b]{0.49\textwidth}
         \centering
         \includegraphics[width=\textwidth]{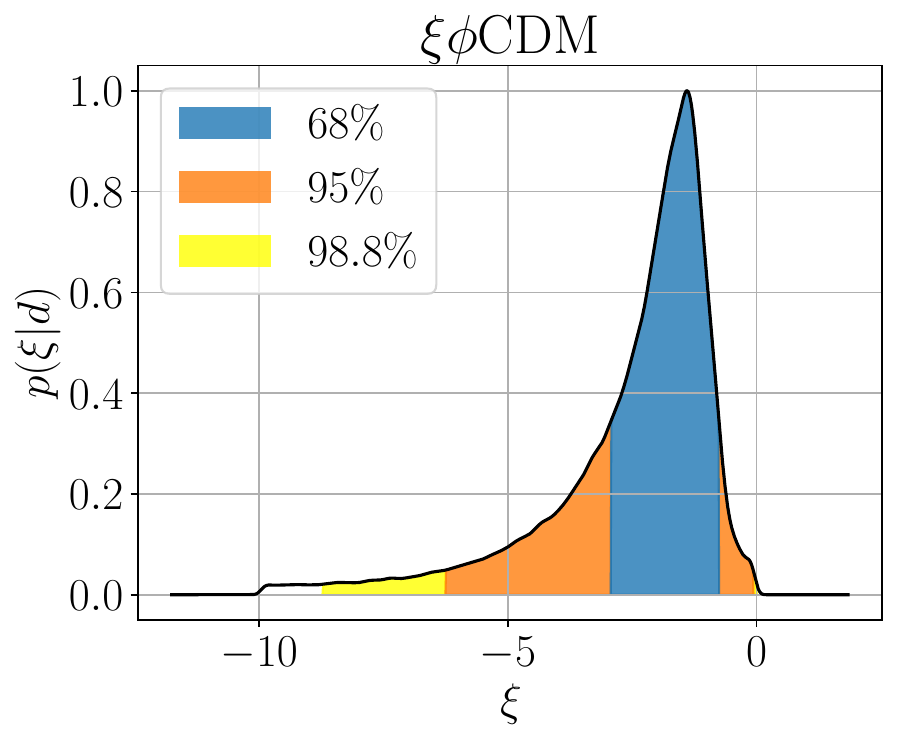}
     \end{subfigure}
     \begin{subfigure}[b]{0.49\textwidth}
         \centering
         \includegraphics[width=\textwidth]{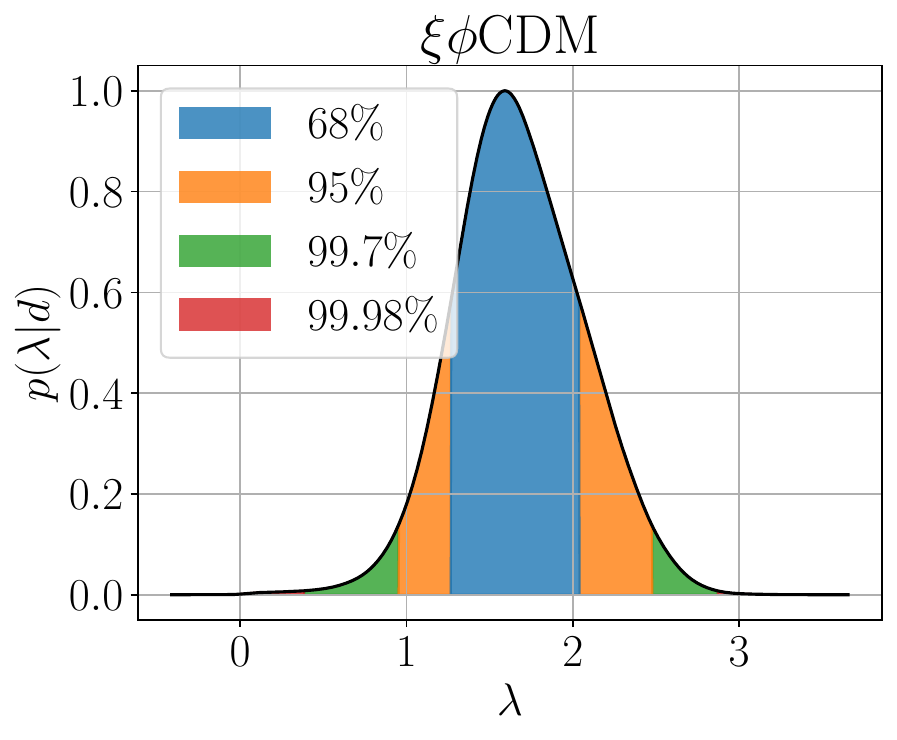}
     \end{subfigure}
    \begin{subfigure}[b]{0.49\textwidth}
         \centering
         \includegraphics[width=\textwidth]{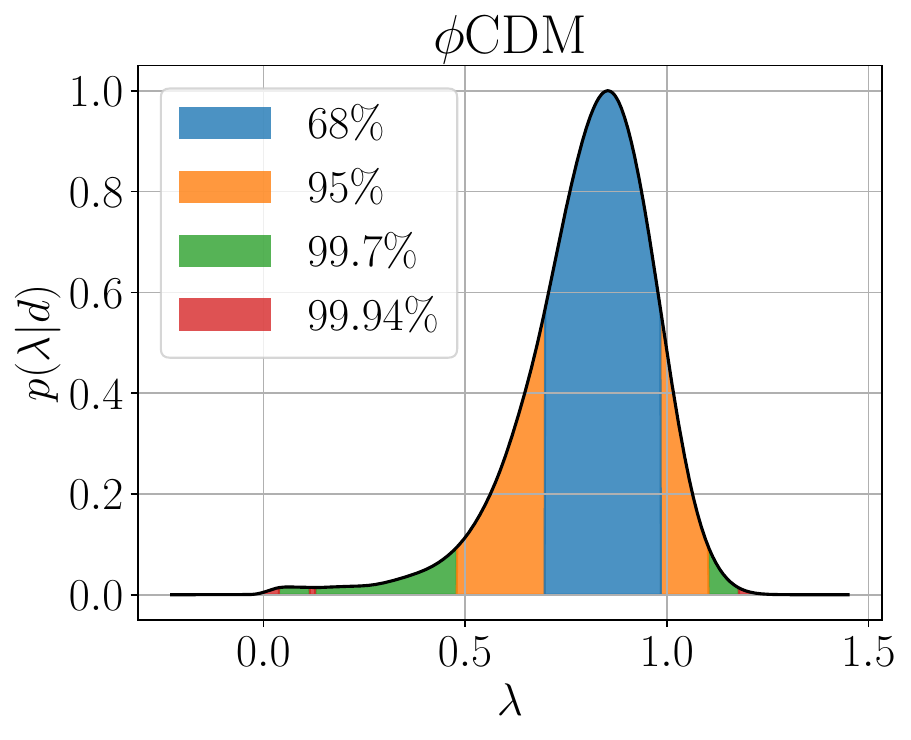}
     \end{subfigure}
    \begin{subfigure}[b]{0.49\textwidth}
         \centering
         \includegraphics[width=\textwidth]{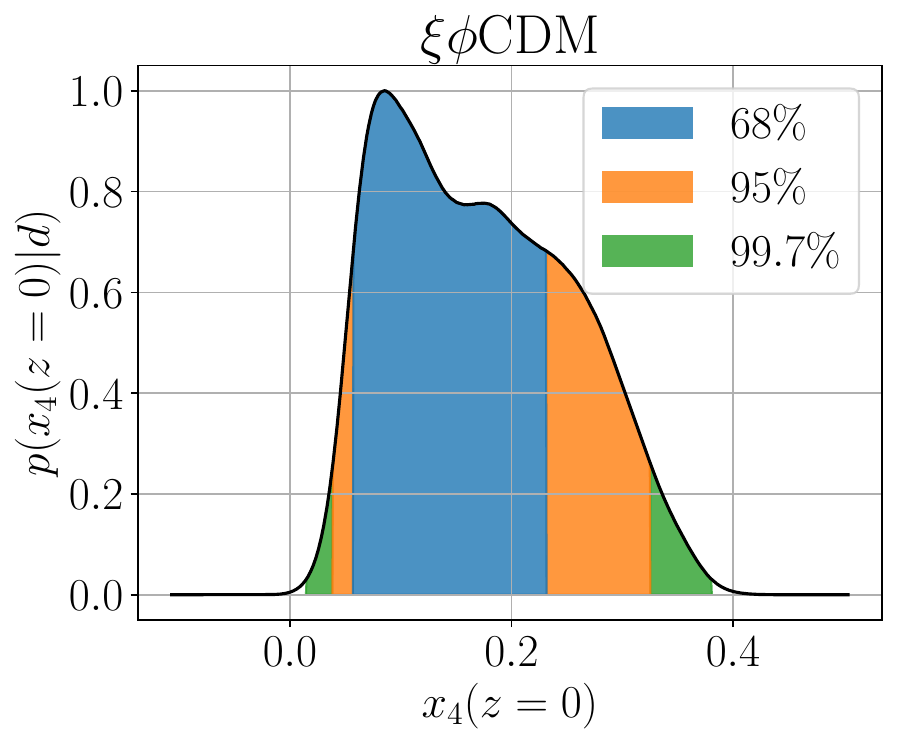}
     \end{subfigure}
     \caption{Posterior distribution of $\xi$ (top left), $\lambda$ in $\xi\phi$CDM (top right), $\lambda$ in $\phi$CDM (bottom left), and the total field excursion $x_4(z=0)=\frac{\Delta \phi}{\m}$ in $\xi\phi$CDM (bottom right) using CMB+DESI+DESY5, with the 68\% (blue), 95\% (orange), and 99.7\% (green) credible intervals. For $p(\xi|d)$ we show the largest credible interval such that $\xi<0$ in yellow. For $p(\lambda|d)$, both in $\xi\phi$CDM and $\phi$CDM, we show the largest credible interval such that $\lambda >0$ in red. From $p(x_4(z=0)|d)$ we see that the total field excursion remains sub-Planckian.}
     \label{fig:confidenceintervals}
\end{figure}
In order to be more precise, we utilize the Kernel Density Estimate (KDE) from \texttt{GetDist} to obtain the different credible intervals at which $\xi$ and $\lambda$ are non-zero, \textit{i.e.}, we find the significance at which dark energy is dynamical and non-minimally coupled. The results are shown in Fig. \ref{fig:confidenceintervals}. For the non-minimal coupling, we find that $\xi<0$ at $98.8\%$ C.L., making it larger than $2\sigma$ but not quite $3\sigma$. Regarding the slope of the exponential, we find that $\lambda > 0$ at $99.98\%$ and $99.94\%$, for $\xi\phi$CDM and $\phi$CDM, respectively. This means that in both models $\lambda$ is non-zero at more than $3\sigma$. 

\begin{figure}[h]
     \centering
     \begin{subfigure}[b]{0.49\textwidth}
         \centering
         \includegraphics[width=\textwidth]{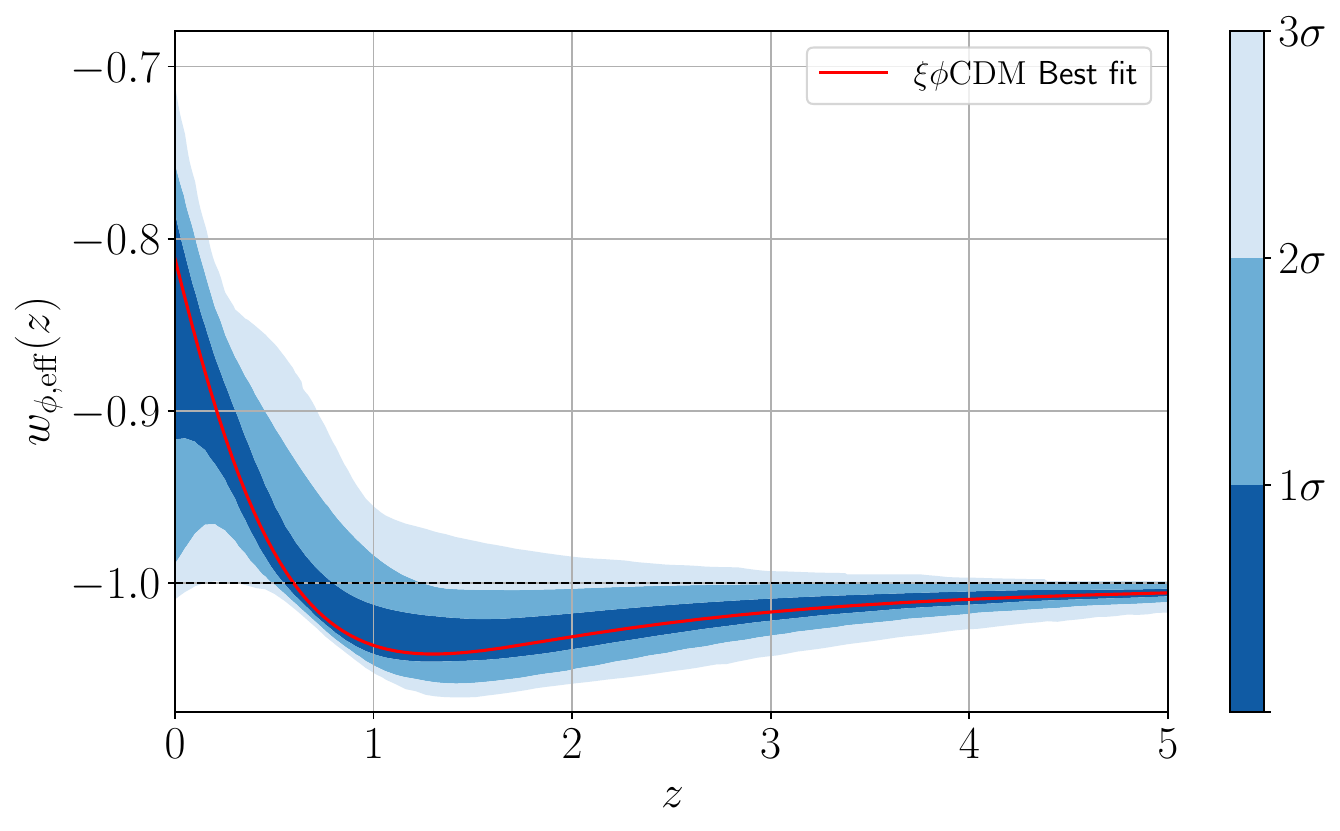}
     \end{subfigure}
     \begin{subfigure}[b]{0.49\textwidth}
         \centering
         \includegraphics[width=\textwidth]{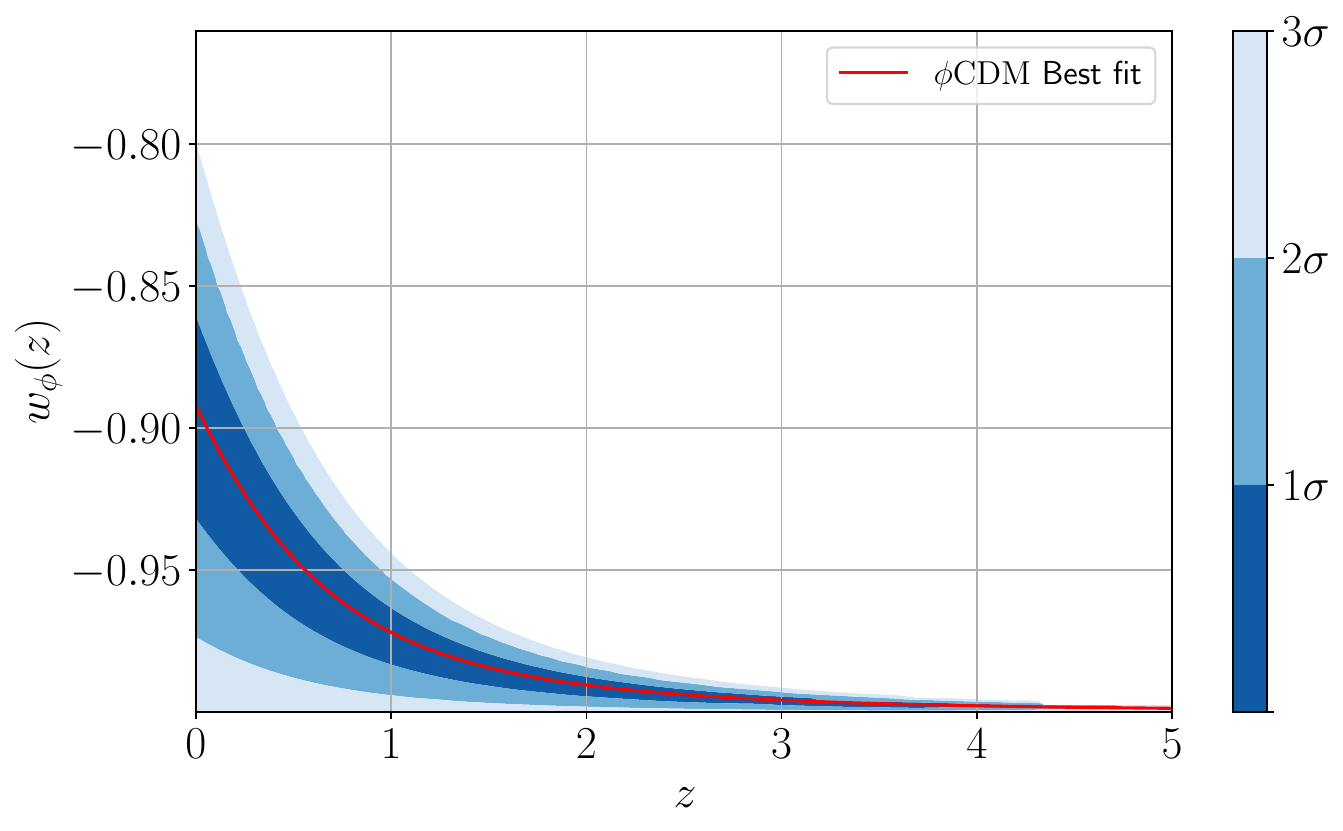}
     \end{subfigure}
     \caption{Posterior distributions of the effective barotropic parameter of quintessence as a function of redshift for $\xi\phi$CDM (left) and $\phi$CDM (right), given the posterior distributions of the model parameters. We show the 68\%, 95\%, and 99.7\% confidence intervals in progressively lighter shades of blue and the best-fit barotropic parameter in red.}
     \label{fig:wcontours}
\end{figure}

Finally, we plot the posterior distribution of the effective barotropic parameter of the field, for both $\xi\phi$CDM and $\phi$CDM, as a function of redshift in Fig.~\ref{fig:wcontours}. Functional posterior distributions were generated using the publicly available \texttt{Python} package \texttt{fgivenx}~\cite{fgivenx}, utilizing the same samples as those in the \texttt{MCEvidence} analysis. For the non-minimal coupling case, we find that the $\Lambda$CDM line $w=-1$ lies outside of the $2\sigma$ confidence band, indicating a preference for a phantom crossing at $\lesssim 3\sigma$ C.L.. As for the minimal coupling case, the data prefers a barotropic parameter $w\neq -1$ at more than $2\sigma$ but less than $3\sigma$. However, above this confidence level, the data is unable to distinguish between dynamical dark energy and a cosmological constant. Even though we found that strictly speaking $\lambda > 0$ at more than $3\sigma$, the time evolution of $w_{\phi}$ also depends on the field dynamics. Indeed, our numerical analysis reveals that to have $w_{\phi,{\rm eff}}(z=0)>-0.99$ one needs $\lambda > 0.25$, which lies in the $98\%$ C.L., consistent with Fig.~\ref{fig:wcontours}. However, a smaller value like $\lambda = 0.08$ gives $w_{\phi,{\rm eff}}(z=0)=-0.9992$, extremely close to $\Lambda$CDM.

\subsection{Palatini vs Metric} \label{eqaouosdbfadfdawer}
In this section, we assess the question of whether the data has anything to say about the degrees of freedom of the theory of gravity. To do so, we repeat the entire MCMC analysis pipeline in the metric formalism, labeled $\Tilde{\xi}\phi$CDM for convenience.
 \begin{table}[H]
    \centering
    \small
    \begin{tabular}{|>{\centering\arraybackslash}p{1.0cm}|>{\centering\arraybackslash}p{4.1cm}|>{\centering\arraybackslash}p{4.1cm}|>{\centering\arraybackslash}p{4.1cm}|}
        \hline
         Params  & Metric $\tilde{\xi}\phi$CDM                              & Palatini $\xi\phi$CDM \\
        \hline\hline
        $\xi$       &           $-3.68(-2.05)^{+3.1}_{-0.97}$                 & $-2.50(-1.41)^{+1.7}_{-0.42}$\\
        
        $\lambda$   &    $1.72(1.61)^{+0.33}_{-0.47}$ & $1.68(1.95)^{+0.36}_{-0.41}$\\
        
        $\Omega_{\rm m}$ &   $0.3163(0.3174)\pm 0.0056$    & $0.3179(0.3193)\pm0.0059$\\
        
        $H_0$            &  $66.95(66.85)\pm 0.56$       & $66.82(66.70)\pm0.58$\\
        
        $\Omega_{\rm b}$ &  $0.05019(0.05029)\pm 0.00084$  & $0.05036(0.05049)\pm0.00086$\\
        \hline
    \end{tabular}
        \caption{CMB+BAO+SN 68$\%$ ($1\sigma$) credible intervals and best-fit values (in parentheses) for the parameters in the metric $\tilde{\xi}\phi$CDM and Palatini $\xi\phi$CDM formalisms.}
        \label{table:parameterconfidencemetric}
\end{table}
\begin{table}[H]
    \centering
    \begin{tabular}{|l|l|c|c|c|c|}
        \hline
         & Model & Planck & DESI DR2 & DESY5 & Total \\
        \hline\hline
       \multirow{2}{*}{$\Delta \chi^2$} &$\xi\phi$CDM & 0.46 & -3.41 & -11.70 & -14.66\\
        &$\tilde{\xi}\phi$CDM & 0.83 & -2.96 & -10.82 & -12.95\\
        \hline
        \multirow{2}{*}{$\log B$} &$\xi\phi$CDM & - & - & - & 5.52\\
        &$\tilde{\xi}\phi$CDM & - & - & - & 4.78\\
        \hline
    \end{tabular}
        \caption{Change in $\chi^2$ (first two rows) and $\log B$ (last two rows) for the metric $\tilde{\xi}\phi$CDM and Palatini $\xi\phi$CDM formalisms, relative to $\Lambda$CDM, evaluated at the CMB+BAO+SN best-fit parameters. The contribution from each dataset is shown in the corresponding column, with the total reported in the last column. A negative (positive) $\Delta \chi^2$ corresponds to an improvement (worsening) in fit. A positive (negative) $ \log B$ implies evidence in favor (against) of the model over $\Lambda$CDM.}
        \label{table:statisticssummarymetric}
\end{table}
\begin{figure}[h]
         \centering
         \includegraphics[width=1\textwidth]{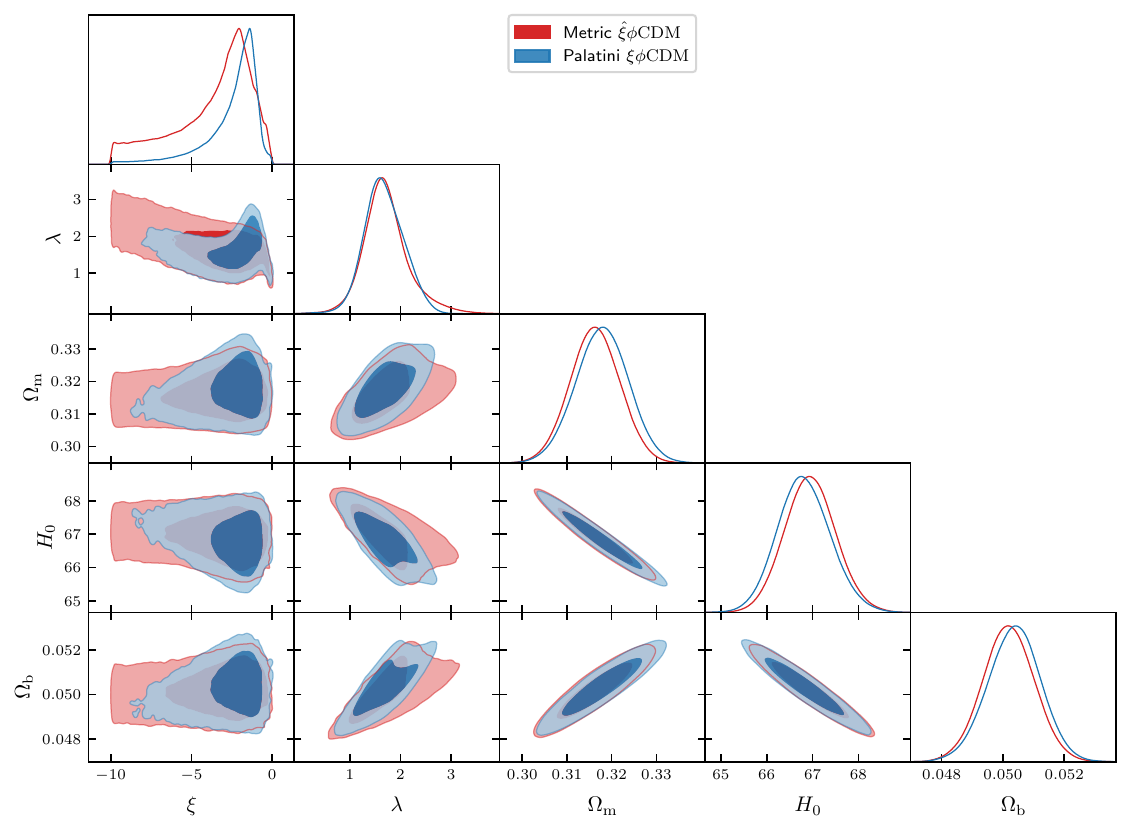}
     \caption{Parameter posteriors for in the Palatini $\xi\phi$CDM  (blue) and metric $\tilde{\xi}\phi$CDM  (red) formalisms using CMB+DESI+DESY5. The darker (lighter) shaded regions represent the 68$\%$ (95$\%$) credible intervals.}
     \label{fig:cornermetric}
\end{figure}
\begin{figure}[h]
         \centering
         \includegraphics[width=0.8\textwidth]{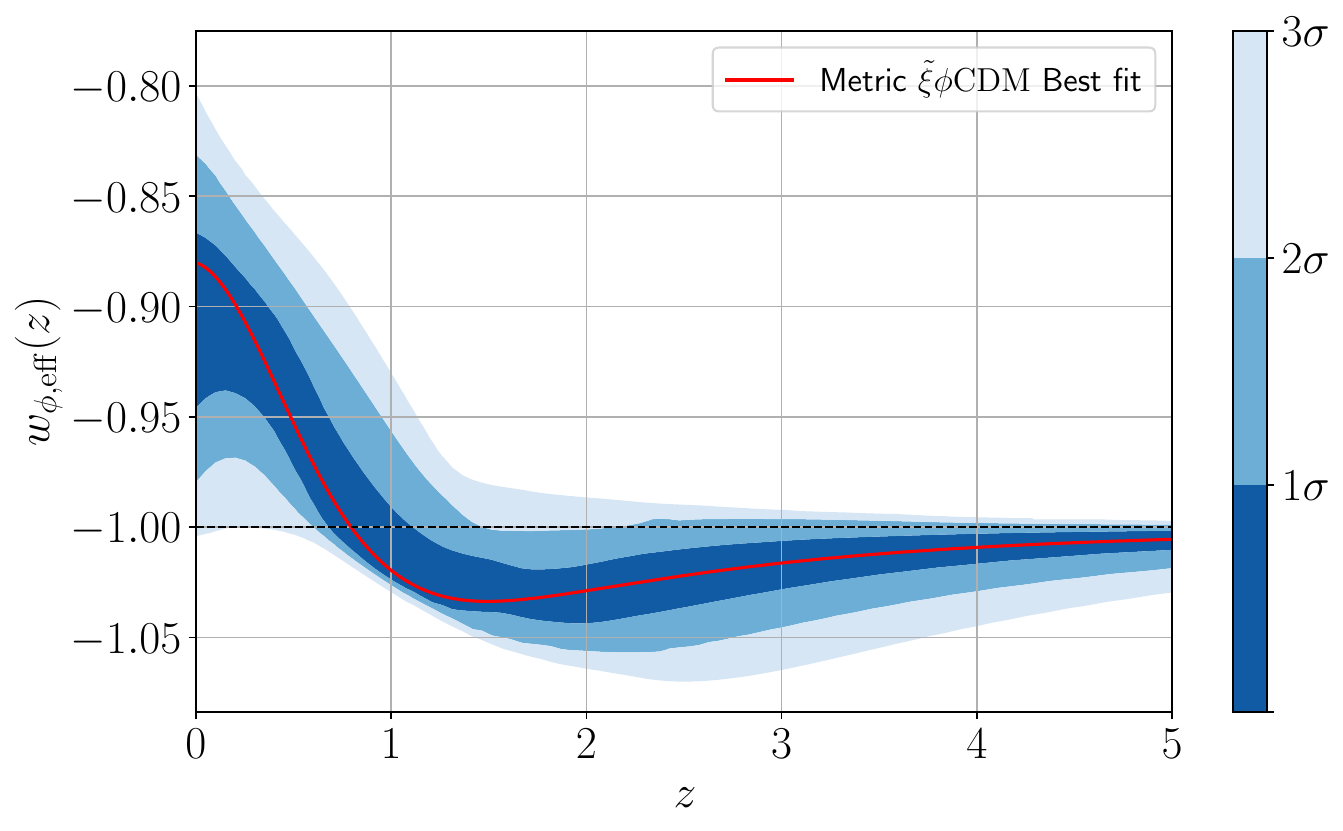}
     \caption{Posterior distributions of the effective barotropic parameter of quintessence as a function of redshift for metric $\tilde{\xi}\phi$CMD, given the posterior distributions of the model parameters. We show the 68\%, 95\%, and 99.7\% confidence intervals in progressively lighter shades of blue and the best-fit barotropic parameter in red.}
     \label{fig:wcontoursmetric}
\end{figure}
In Table~\ref{table:parameterconfidencemetric} we report the CMB+BAO+SN 68$\%$ confidence intervals for the parameters of $\tilde{\xi}\phi$CDM along with its best-fit values. For the convenience of the reader, we also repeat the $\xi\phi$CDM values. Just like in the Palatini formalism, the late-time accelerating attractor exists for all values of $\lambda$ as long as $\xi<0$, something that is reflected in the best-fit value and confidence interval of $\xi$. In Table~\ref{table:statisticssummarymetric} we report the goodness of fit of $\tilde{\xi}\phi$CDM relative to $\Lambda$CDM in terms of the difference in $\chi^2$, for each dataset as well as the total, given the best-fit parameters. We also report the Bayes factor. We again provide the Palatini formalism values for convenience. The fit of $\Tilde{\xi}\phi$CDM to all data sets is slightly worse than that of $\xi\phi$CDM, although the difference between formalisms is not large, amounting to a total $\Delta \chi^2 = -1.17$ in favor of the Palatini formalism. This is reflected in the Bayes factor, obtaining moderate evidence of $\tilde{\xi}\phi$CDM over $\Lambda$CDM, albeit close to strong, according to the Jeffrey's scale.

Lastly, we plot the posterior distribution of the effective barotropic parameter of the field in Fig.~\ref{fig:wcontoursmetric}. As in the Palatini formalism, we find a preference for a phantom crossing at $\lesssim 3\sigma$ C.L., although the significance seems slightly smaller. Furthermore, the $1\sigma$ confidence band covers a lower value for $w_{\phi,{\rm eff}}$ at present $z=0$.

\section{Conclusions} \label{ref:conclusions}
We have explored the role of a scalar field, non-minimally coupled to gravity, in describing a range of cosmological observations and tested the importance of the non-minimal coupling. Our main focus is on the Palatini formalism of gravity, where the connection is taken to be an independent gravitational field. This is contrary to the metric formalism, with the connection being fixed to its standard Levi-Civita form, leading to different dynamics. Among other modifications, for an FLRW metric, the Palatini Friedmann and Raychaudhuri equations, as well as the Klein-Gordon equation, feature additional terms with respect to their metric counterparts. This is also apparent by writing the field equations as a closed dynamical system, after choosing the quintessence potential to be an exponential.

Most importantly, the non-minimal coupling (in both formalisms) allows for the barotropic parameter of the field to cross the phantom divide and become smaller than $-1$ in the redshift range $0.5\lesssim z \lesssim 4$. This is, however, a transitory phase, as our phase space analysis reveals a late-time de Sitter attractor, the existence of which is guaranteed as long as the non-minimal coupling constant is negative. Both formalisms share the same coordinates in phase space for this attractor, although, again, the details of the dynamics are different. 

In this work, we have for the first time analyzed a model of Palatini non-minimally coupled quintessence in the light of state-of-the-art cosmological data, and tested whether the data has a preference for the formalism of gravity. We compare the theory against the most recent datasets from three experiments that probe observables in the Universe at different epochs. In particular, we use the summary statistics from the Cosmic Microwave Background observations, obtained from the final Planck data release, distance moduli from the Dark Energy Survey Year 5, and the second data release of the geometrical measurements from the Dark Energy Spectroscopic Instrument. Our analysis also includes the standard $\Lambda$CDM model of cosmology, minimally coupled quintessence, and non-minimally coupled quintessence in the metric formalism.  

Quantifying the goodness of fit with the difference in $\chi^2$, we find an improvement of nearly $15$ when compared to $\Lambda$CDM. The breakdown of $\chi^2$ indicates the improvement is mainly driven by the SNe and BAO data, while the CMB is addressed similarly by both models, non-minimally coupled or not, since the scalar field evolution at early times ensures $\Lambda$CDM behavior. The Bayes factor with respect to $\Lambda$CDM is 5.52, providing strong evidence for the new model, according to the Jeffrey's scale. In the metric formalism, the improvement in fit (and so the Bayes factor) with respect to $\Lambda$CDM from the non-minimal coupling is reduced marginally compared to its Palatini counterpart.  

In two steps, the improvement can be explained as follows: firstly, we find that $\lambda > 0$ at 99.94\% C.L. for a minimally coupled scalar field. This means that the three datasets, when combined, prefer an evolving equation-of-state for dark energy at $>3\sigma$. Furthermore, for the non-minimally coupled field, we find that $\xi<0$ at 98.8\% C.L., \textit{i.e.}, a 2-3$\sigma$ preference for the coupling term. Secondly, from a qualitative point of view, the observations prefer a phantom crossing in the barotropic parameter of dark energy between $z=0.5-4$ and a steep rise to $w>-1$ at $z<0.5$. This behavior changes impose a crossing in the distance modulus with respect to the $\Lambda$CDM model and a dip in the comoving and Hubble distance near $z=0.4$ that improves the fit to both datasets. 

Our results demonstrate, for the first time, that a non-minimally coupled scalar field in the Palatini formalism significantly improves the fit to the CMB, BAO, and SNe data in a joint analysis, when compared to $\Lambda$CDM. The improvement rules out any statistical uncertainties and leaves room only for the systematics in the SNe or BAO measurements as an example that counters the beyond standard model physics. Our results also demonstrate a marginal improvement with respect to the analogous theory in the metric formalism. 

The evidence for a non-minimal coupling also raises challenges for the future. Indeed, its presence leads to a time evolution in the effective gravitational constant and to fifth forces~\cite{Carroll:1998zi}, both tightly constrained by experiments~\cite{Adelberger:2003zx, Will:2014kxa}. Regarding the first, a time-varying gravitational constant changes the Poisson equation for the gravitational potential, thereby affecting the growth of structure. As for the latter, the Parametrized Post-Newtonian parameters are strongly bounded by Solar System experiments, such as the Cassini data~\cite{Bertotti:2003rm}. Although we find a sub-Planckian field displacement in our model, these issues should be addressed in detail. We leave that for future work.

\newpage

\section*{Acknowledgements}
We thank Kushal Lodha for insightful discussions. SSL and DKH would like to thank the Indo-French Centre for the Promotion of Advanced Research (IFCPAR/CEFIPRA) for support of the proposal 6704-4 titled `Testing flavors of the early universe beyond vanilla models with cosmological observations’ under the Collaborative Scientific Research Programme. AK was supported by the Estonian Research Council grants PSG761, RVTT3, RVTT7 and by the Center of Excellence program TK202. Computational portions of this work were carried out on the Kamet high performance computing cluster at the Institute of Mathematical Sciences (IMSc), Chennai, maintained and supported by the High-Performance Computing Center of IMSc.

\bibliography{bibliography}

\end{document}